\documentclass{aa}  
\usepackage{hyperref}
\usepackage{graphicx}
\usepackage{txfonts}
\usepackage{adjustbox}
\usepackage{lineno}
\begin{document} 

\title{A multifrequency characterization \\ of the extragalactic hard X-ray sky}
\subtitle{Presenting the 2$^{nd}$ release of the Turin-SyCAT}

   \author{M. Kosiba\inst{1, 2}
          \and
          H. A. Pe\~{n}a-Herazo\inst{3}
          \and
          F. Massaro\inst{2, 4, 5}
          \and
          N. Masetti\inst{6, 7}
          \and
          A. Paggi\inst{2, 3, 4}
          \and  \\
          V. Chavushyan\inst{8}
          \and
          E. Bottacini\inst{9, 10}
          \and
          N. Werner\inst{1}
          }

    \institute{Department of Theoretical Physics and Astrophysics, Faculty of Science, Masaryk                    University, Kotl\'a\v rsk\'a 2, Brno, 611 37, Czech \\
               Republic
        \and
            Dipartimento di Fisica, Universit\`{a} degli Studi di Torino, via Pietro Giuria 1, I-10125 Torino, Italy.
        \and
            East Asian Observatory, 660 North A'oh{\=o}k{\=u} Place, Hilo, Hawaii 96720, USA.
        \and
            Istituto Nazionale di Fisica Nucleare, Sezione di Torino, I-10125 Torino, Italy
        \and
            INAF–Osservatorio Astrofisico di Torino, via Osservatorio 20, I-10025 Pino Torinese, Italy
        \and
            INAF - Osservatorio di Astrofisica e Scienza dello Spazio, via Piero Gobetti 101, 40129 Bologna, Italy
        \and
            Departamento de Ciencias F\'isicas, Universidad Andr\'es Bello, Fern\'andez Concha 700, Las Condes, Santiago, Chile
        \and
            Instituto Nacional de Astrofísica, \'Optica y Electr\'onica, Apartado Postal 51-216, 72000 Puebla, M\'exico.
        \and 
            Dipartimento di Fisica e Astronomia G. Galilei, Univerist\`a di Padova, Padova, Italy.
        \and
            Eureka Scientific, 2452 Delmer Street Suite 100, Oakland, CA 94602-3017, USA.
        }

\date{Received Month Day, Year; accepted Month Day, Year}

\abstract
{Nowadays we know that the origin of the Cosmic X-ray Background (CXB) is due to the integrated emission of nearby active galactic nuclei. Thus to obtain a precise estimate of the contribution of different source classes to the CXB it is crucial to have a full characterization of the hard X-ray sky.}
{We present a multifrequency analysis of all sources listed in the 3$^{rd}$ release of the Palermo \textit{Swift}-BAT hard X-ray catalog (3PBC) with the goal of (i) identifying and classifying the largest number of sources adopting multifrequency criteria, with particular emphasis on extragalactic populations and (ii) extracting sources belonging to the class of Seyfert galaxies to present here the release of the 2$^{nd}$ version of the Turin-SyCAT.}
{We outline a classification scheme based on radio, infrared and optical criteria that allows us to distinguish between unidentified and unclassified hard X-ray sources, as well as to classify those sources belonging to the Galactic and the extragalactic populations.}
{Our revised version of the 3PBC lists 1176 classified, 820 extragalactic, and 356 Galactic ones, 199 unclassified, and 218 unidentified sources. According to our analysis, the hard X-ray sky is mainly populated by Seyfert galaxies and blazars. For the blazar population, we report trends between the hard X-ray and the gamma-ray emissions since a large fraction of them have also a counterpart detected by the \textit{Fermi} satellite. These trends are all in agreement with the expectations of inverse Compton models widely adopted to explain the blazar broadband emission. For the Seyfert galaxies, we present the $2^{nd}$ version of the Turin-SyCAT including a total of 633 Seyfert galaxies, with 282 new sources corresponding to an increasement by $\sim$80\,\% with respect to the previous release. Comparing the hard X-ray and the infrared emissions of Seyfert galaxies we confirm, that there is no clear difference between the flux distribution of the infrared-to-hard X-ray flux ratio of Seyfert galaxies Type 1 and Type 2. However, there is a significant trend between the mid-IR flux and hard X-ray flux, confirming previous statistical results in the literature.}
{We provide two catalog tables. The first is the revised version of the 3PBC catalog based on our multifrequency analyses. The second catalog table is a release of the second version of the Turin-SyCAT catalog. Finally, we highlight that the SWIFT archive has already extensive soft X-ray data available to search for potential counterparts of unidentified hard X-ray sources. All these datasets will be reduced and analyzed in a forthcoming analysis to determine the precise position of low energy counterparts in the 0.5 -- 10 keV energy range for 3PBC sources that can be targets of future optical spectroscopic campaigns, necessary to obtain their precise classification.}

\keywords{catalogs -- methods: data analyses -- X-rays: general}
\maketitle


\section{Introduction}
The Cosmic X-ray background (CXB) was discovered when the earliest X-ray astronomical rocket experiments were carried out \citep[see e.g.][]{Giacconi1962}. It appeared as a diffuse component of X-ray radiation distributed in all directions. In the last decades, after its discovery, several different scenarios were proposed to interpret its origin, such as considering spanning new types of faint discrete X-ray sources whose integrated emission could be responsible for the CXB (e.g. \citep{Gilli1999}, \cite{Gilli2001}) up to diffuse radiative processes occurring in the intergalactic space, as for example exotic emission from dark matter particle decay \citep[see e.g.][]{Abazajian2001}. However, the solution to this puzzle arose thanks to deep images obtained first with the ROentgen SATellite \textit{ROSAT} \citep{Hasinger1999}, collected in the early ninety's, and more recently with {\it Chandra} X-ray telescope \citep{Weisskopf2000}, all revealing that about 80\,\% of CXB is resolved \citep{Hasinger1998}, between 0.5\,keV and\,2 keV, as suggested by \citep{Cavaliere1976}. 
At hard X-ray energies the fraction of resolved CXB by {\em Swift} and {\em INTEGRAL} is of 2\% \citep{Bottacini2012} and by {\em NuSTAR} is of 35\% \citep{Harrison2016}.
Thus, the origin of the CXB is nowadays established to be mainly due to the high energy emission of the extragalactic discrete sources, whose large fraction belongs to different classes of active galactic nuclei (AGNs) \citep{Gilli2007}.

The first survey in the hard X-ray band was carried by the \textit{UHURU} satellite \citep[a.k.a SAS-1][]{Giacconi1971}. Since the discovery of the CXB, many surveys were performed in the soft and hard X-ray bands, including \citet{Forman1978} who produced a catalog of 339 X-ray sources observed by the \textit{UHURU} satellite in the 2--20~keV energy band. \citet{Levine1984}, using the X-ray and Gamma-ray detector HEAO-A4 on board the \textit{HEAO~1} satellite \citep{Rothschild1979} presented an all-sky survey in 13--180~keV range detecting 77 new sources. The hard X-ray component of the CXB radiation, observable between 3\,keV and up to 300\,keV, shows a distinct peak at $\sim$\,30 keV \citep{Gruber1999} being extremely uniform across the sky with the only exception of an overdensity along the Galactic plane \citep{Valinia1998, Revnivtsev2006, Krivonos2007a} and it is again strictly connected with AGN population emitting in the hard X-rays \citep{Frontera2007}.


The currently flying satellites as the INTErnational Gamma-Ray Astrophysics Laboratory \textit{INTEGRAL} \citep{Winkler2003}, with its Imager on Board the \textit{INTEGRAL} Satellite \textit{IBIS} \citep{Ubertini2003} and Neil Gehrels \textit{Swift} Observatory \citep{Gehrels2004} with its Burst Alert Telescope \citep[BAT][]{Barthelmy2004} on board, carrying measurements in the hard X-rays significantly improved our understanding of the origin of the CXB and refined its measurement. The observed spectrum of the CXB is currently well described by the standard population synthesis model of AGNs, including the fraction of Compton-thick AGNs and the reflection strengths from the accretion disk and torus based on the luminosity- and redshift-dependent unified scheme \citep{Ajello2008, Ueda2014}. This is shown also in recent results achieved thanks to the NuSTAR observations in the 3– 20 keV band \citep{Krivonos2021b}. 
This has been also possible thanks to improvements achieved in the preparation of hard X-ray source catalogues \citep[see e.g.][]{Markwardt2005, Beckmann2006, Churazov2007, Krivonos2007b, Sazonov2007, Tueller2008, Cusumano2010, Bottacini2012, Bird2016, Mereminskiy2016, Krivonos2017, Oh2018, Krivonos2021, Krivonos2022} and the association of hard X-ray sources with their low energy counterparts \citep[e.g.][]{Malizia2010, Koss2019, Bar2019, Smith2020} and their optical spectroscopy follow-ups \citep[e.g.][]{Masetti2006a, Masetti2006b, Masetti2008, Masetti2012, Masetti2013, Parisi2014, Rojas2017, Marchesini2019}.

There are three major catalogues built on observations collected in the last decade with two major space missions still active: (i) the Palermo \textit{Swift}-BAT hard X-ray catalogue \citep{Cusumano2010} based on 54 months of the Swift-BAT operation, currently updated to its 3$^{rd}$ release and with a 4$^{th}$ release ongoing\footnote{\href{https://www.ssdc.asi.it/bat54/}{https://www.ssdc.asi.it/bat54/}}, (ii) the \textit{Swift}-BAT all-sky hard X-ray survey, that published the 105 month \textit{Swift}-BAT catalogue \citep[see e.g.][]{Oh2018}, and (iii) the \textit{INTEGRAL} IBIS catalogue in the energy range 17-100 keV \citep{Bird2016}, performed using the \textit{INTEGRAL} Soft $\gamma$-ray Imager (ISGRI) \citep{Lebrun2003}, the low energy CdTe $\gamma$-ray detector on (\textit{IBIS}) telescope \citep{Ubertini2003}.

Here we focus on the investigation of the 3$^{rd}$ release of the Palermo \textit{Swift}-BAT hard X-ray catalog (hereinafter 3PBC), with particular emphasis on extragalactic sources, since the release of the next version is currently ongoing, thus results provided by our analysis could be used therein. The 3PBC is based on the data reduction and detection algorithms of the first Palermo {\it Swift}-BAT Catalog hard X-ray catalog \citep{Segreto2010, Cusumano2010}. The 3PBC is available only online\footnote{\href{http://bat.ifc.inaf.it/bat\_catalog\_web/66m\_bat\_catalog.html}{http://bat.ifc.inaf.it/bat\_catalog\_web/66m\_bat\_catalog.html}} thus we point for the reference the publication of its 2$^{nd}$ release, the 2PBC \citep{Cusumano2010}. The 2PBC provides data in three energy bands, namely: 15 -- 30 keV, 15 -- 70 keV, 15 -- 150 keV for a total of 1256 sources above 4.8 $\sigma$ level of significance, where 1079 hard X-ray sources have an assigned soft X-ray counterpart, while the remaining 177 are still unassociated. The total source number increased in 3PBC to 1593 when considering a signal-to-noise ratio above 3.8, which is the catalog release we analyzed here. Please note, that only three 3PBC sources are detected at a signal-to-noise ratio lower than 5.
The 3PBC catalogue covers 90\,\% of the sky down to a flux limit of 1.1\,$\times$ 10$^{-11}$\,erg\,cm$^{-2}$\,s$^{-1}$, decreasing to $\sim$50\,\% when decreasing the flux limit to 0.9\,$\times$ 10$^{-11}$\,erg\,cm$^{-2}$\,s$^{-1}$. 

First, we verified source classification for all associated counterparts listed in the 3PBC, adopting a multifrequency approach. This analysis was corroborated by checking if additional studies, available in the literature and carried out after the last 3PBC release, allowed us to obtain a more complete overview of source populations emitting in the hard X-rays. Then, our final goal was to explore in detail those extragalactic sources being identified as Seyfert galaxies \citep{Antonucci1985} to (i) release the 2$^{nd}$ version of the Turin-SyCAT \citep{Herazo2022} and thus (ii) refine our statistical analysis on the correlation found between the infrared (IR) and the hard X-ray fluxes for this extragalactic population. Additionally, we also aim at investigating possible connections between the hard X-ray and the gamma-ray emission in those blazars detected by Fermi-LAT. It is worth noting that given our final aim, the classification task performed on the Galactic sources is mainly devoted to excluding them from the final sample of new Seyfert galaxies.

The present work will be also relevant for the association of hard X-ray sources with their low energy counterpart, which will be included in the next releases of hard X-ray catalogs. In addition, we highlight that the proposed investigation will also provide a more complete overview of those sources, that lack an assigned low energy counterpart as still unidentified. 

Finally, we remark that the reason underlying the choice of working with the 3PBC rather than subsequent versions of hard X-ray catalogs is mainly motivated by the opportunity of having more multifrequency information available in the literature. However a comparison with other recent catalogues as: the 105 month \textit{Swift}-BAT catalog\footnote{\href{https://heasarc.gsfc.nasa.gov/W3Browse/swift/swbat105m.html}{https://heasarc.gsfc.nasa.gov/W3Browse/swift/swbat105m.html}} \citep{Oh2018} and the \textit{INTEGRAL} hard X-ray catalogue \citep{Bird2016} are also included in the present analysis.

The structure of the paper is outlined as follows: in \hyperref[sec:section2]{Section~2}, we described various catalogs and surveys used to search for multifrequency information related to high and low energy counterparts of hard X-ray sources; in \hyperref[sec:section3]{Section~3}, we present our multifrequency classification scheme adopted to label source counterparts. Then \hyperref[sec:section4]{Section~4} focuses on the main results of the characterization of the extragalactic hard X-ray sources while \hyperref[sec:section5]{Section~5} is entirely devoted to the second release of the Turin-SyCAT catalog and the statistical analysis for the IR -- hard X-ray connection. 
Finally, our summary, conclusions, and future perspectives are given in \hyperref[sec:section6]{Section~6}. A comparison between our classification analysis and the previous one of the 3PBC is then reported in Appendix \ref{app:3PBC_reclassification}.

We used cgs units unless stated otherwise. We adopted $\Lambda$CDM cosmology with $\Omega_M = 0.286$, and Hubble constant $H_{0} = 69.6$\,km\,$s^{-1}\,Mpc^{-1}$ \citep{Bennett2014} to compute cosmological corrections, the same used for the 1$^{sth}$ release of the Turin-SyCAT \citep{Herazo2022}. WISE magnitudes are in the Vega system and are not corrected for the Galactic extinction. As shown in our previous analyses \citep{D'Abrusco2014, Massaro2016, D'Abrusco2019}, such correction affects mainly the magnitude at 3.4 $\mu$\,m for sources lying at low Galactic latitudes (i.e., $|b| <20^{\circ}$ ), and it ranges between 2\,\% and 5\,\% of their magnitude values, thus not significantly affect our results. We indicate the WISE magnitudes at 3.4, 4.6, 12, and 22 $\mu$\,m as W1, W2, W3, and W4, respectively. For all WISE magnitudes of sources flagged as extended in the AllWISE catalog (i.e., extended flag ``ext\_flg'' greater than 0) we used values measured in the elliptical aperture. Sloan Digital Sky Survey (SDSS) \citep{Blanton2017, Abdurro2021} and Panoramic Survey Telescope \& Rapid Response System (Pan-STARRS) \citep{Chambers2016} magnitudes are in the AB system. Given the large number of acronyms used here, mostly due to different classifications and telescopes used, we summarized them in Table \ref{table:Table of Acronyms}.
\begin{table}[h]
\caption{Table of Acronyms used in the text.}           
\label{table:Table of Acronyms}      
\begin{tabular}{cc}
 \hline \hline
  Acronym & Meaning \\
\hline
\hline
ATNF    & australian telescope national facility \\
CXB     & cosmic X-ray background \\
AGN     & active galactic nuclei \\
BLL     & BL-Lac object \\
BZG     & Galaxy Dominated Blazars \\
BZU     & blazar of uncertain type \\
CV      & cataclysmic variable \\
FSRQ    & flat spectrum radio quasar \\
HERG    & high excitation radio galaxy \\
LERG    & low excitation radio galaxy \\
LINER   & Low-ionization nuclear \\
        & emission-line region galaxy \\
NOV     & novae  \\
PN      & planetary nebulae \\
PSR     & pulsar \\
QSO     & quasi stellar objects \\
RDG     & radio galaxy \\
SNR     & supernovae remnant \\
WD      & white dwarf \\
XBONG   & X-ray bright optically normal galaxy \\
\hline
\end{tabular}
\end{table}

\section{Hunting Counterparts of Hard X-ray Sources: Catalogues and Surveys}
\label{sec:section2}
This section provides a basic overview of all major catalogs used to carry out cross-matching analysis across the whole electromagnetic spectrum. Here we considered several (i) low energy and multifrequency catalogs, listing sources detected in radio, infrared and optical surveys and or based on literature analyses, and (ii) high energy catalogs, based on hard X-rays and $\gamma$-ray surveys.

It is worth noting that the 3PBC catalog is based on a moderate shallow survey thus we expect relatively bright sources in the hard X-rays to be also bright at lower energies, at least for the extragalactic population of 3PBC sources, being mainly constituted by AGNs. This limits the number of catalogs used to perform the cross-matching analysis and we used the same adopted in the original 3PBC analysis. Our analysis has been also augmented by using NED\footnote{\href{https://ned.ipac.caltech.edu/}{https://ned.ipac.caltech.edu/}} and SIMBAD\footnote{\href{http://simbad.cds.unistra.fr/simbad/}{http://simbad.cds.unistra.fr/simbad/}} databases, where we queried all sources having a low energy counterpart listed in the 3PBC before providing a final classification to verify the presence of updated literature information that is not reported in the catalogs adopted for the cross-matching analysis. All catalogs used in the current analysis are listed in Table \ref{table:catalogs_table}.

\subsection{Low energy catalogues for cross-matching analysis}
At low frequencies, from radio to X-ray energies below 10\,keV, we mainly considered:
\begin{enumerate}
\item The Revised Third Cambridge catalog\footnote{\href{https://ned.ipac.caltech.edu/uri/NED::InRefcode/1985PASP...97..932S}{https://ned.ipac.caltech.edu/uri/NED::InRefcode/1985PASP...97..932S}} (3CR, \citep{Spinrad1985}). This catalog provides radio and optical data for 298  extragalactic sources, being the most powerful at low radio frequencies. It includes their positions, magnitudes, classification, and redshifts with only 25 sources being still unidentified \citep{Massaro2013, Maselli2016, Missaglia2021}. More than 90\% of the CR population have available multifrequency observations at radio, infrared, optical, and X-ray energies \citep[see e.g.][]{Massaro2015b, Maselli2016, Stuardi2018}.
The 3CR catalogue was create with a flux density limit S$_{178}$\,$\geq$\,2\,$\times$\,10$^{-26}$\,W\,m\,(Hz)$^{-1}$ at 178 MHz, spanning across the northern hemisphere with declination above -5 degrees. The 3CR catalog has been also augmented by a vast suite of multifrequency observations carried out in the last decades that provides all information necessary to have a completed overview of the source classification \citep{Madrid2006, Privon2008, Massaro2010, Massaro2012c, Kotyla2016, Hilbert2016, Balmaverde2019, Gallardo2021, Balmaverde2021}.

\item The Fourth Cambridge Survey catalog (4C) \footnote{\href{http://astro.vaporia.com/start/fourc.html}{
http://astro.vaporia.com/start/fourc.html}} is based on the radio survey which used the large Cambridge interferometric telescope at the Mullard Radio Astronomy Observatory at frequency 178 Mc\,s$^{-1}$, detecting sources that have flux density S$_{178}$\,$\geq$\,2\,$\times$\,10$^{-26}$\,W\,m\,(Hz)$^{-1}$. It is published in two papers, the first one listing 1219 sources at declination between $+$\,20$^{\circ}$\,and\,$+$\,40$^{\circ}$ \citep{Pilkington1965}, while the second one includes 3624 sources in two declination ranges,  $-$\,07$^{\circ}$\,to\, $+$\,20$^{\circ}$ and $+$\,40$^{\circ}$\,to\, $+$\,80$^{\circ}$ \citep{Gower1967}.

\item The Australia Telescope National Facility (ATNF) \footnote{\href{https://heasarc.gsfc.nasa.gov/W3Browse/all/atnfpulsar.html}{https://heasarc.gsfc.nasa.gov/W3Browse/all/atnfpulsar.html}} Pulsar catalog \citep{Manchester2005} is a complete catalog listing more than 1500 pulsars (PSR). Accretion-powered X-ray PSRs are not included in this catalog, because they have different periods, unstable on short timescales. The catalog is based on the PSR database of 558 PSRs \citep{Taylor1993} which was further supplemented by more recent PSR databases \citep{Manchester2001, Edwards2001} to establish the ATNF PSR catalog. 

\item The Catalog of Galactic Supernovae Remnants (SNRs) \footnote{\href{https://heasarc.gsfc.nasa.gov/W3Browse/all/snrgreen.html}{https://heasarc.gsfc.nasa.gov/W3Browse/all/snrgreen.html}} \citep{Green2017}, which is an updated version of the original catalog of galactic SNRs \citep{Green1984}, currently listing 295 SNRs built on the available results published in literature updated to 2016.

\item The 4$^{th}$ edition of the catalog of High mass X-ray binaries in the Galaxy \footnote{\href{https://heasarc.gsfc.nasa.gov/w3browse/all/hmxbcat.html}{https://heasarc.gsfc.nasa.gov/w3browse/all/hmxbcat.html}} \citep{Liu2006} provides 114 sources, updated with 35 new sources detected, most of them being X-ray binaries having a Be type star or a supergiant star as an optical companion.

\item The 7$^{th}$ edition \footnote{\href{https://heasarc.gsfc.nasa.gov/W3Browse/all/ritterlmxb.html}{https://heasarc.gsfc.nasa.gov/W3Browse/all/ritterlmxb.html}} of the catalog of cataclysmic variables (CVs), low mass X-ray binaries and related objects (original paper \cite{Ritter2003}) lists 1166 cataclysmic variables, 105 low-mass X-ray binaries, and 500 related objects for a total of 1771 sources. The sources are provided with coordinates, apparent magnitudes, orbital parameters, stellar parameters, and other characteristics. The entire catalog is split into three tables provided online.

\item The 4$^{th}$ edition of the catalog of Low mass X-ray binaries in the Galaxy and Magellanic Clouds\footnote{\href{https://heasarc.gsfc.nasa.gov/w3browse/all/hmxbcat.html}{https://heasarc.gsfc.nasa.gov/W3Browse/all/lmxbcat.html}} \citep{Liu2007} contains 187 sources, updated by 44 newly discovered sources.
The companion star of a Low mass X-ray binary is typically a K or M-type dwarf star. Small percentages of the companion stars are G type, red giants, or white dwarfs, and even smaller percentages of companions are A and F type stars.
Sources are provided with their optical counterparts, spectra, X-ray luminosities, system parameters, stellar parameters of the components, and other parameters.

\item The Catalog and Atlas of Cataclysmic Variables\footnote{\href{https://heasarc.gsfc.nasa.gov/W3Browse/all/cvcat.html}{https://heasarc.gsfc.nasa.gov/W3Browse/all/cvcat.html}} (CVcat, \cite{Downes2005}) presented its final release in January 2006 listing 1600 sources. The catalog provides all types of cataclysmic variables like novae, dwarf-novae, nova-like variables, sources classified only as CVs, interacting binary WDs, and possible supernovae. This catalog contains also all objects that have been classified as CVs at some point in the past and are no longer considered to be CVs. Those stars are labeled as NON-CV and are provided also with relevant references.

\item To cross-match the sources with galaxy clusters we used only the Abell catalog of rich galaxy clusters\footnote{\href{https://heasarc.gsfc.nasa.gov/W3Browse/all/abell.html}{https://heasarc.gsfc.nasa.gov/W3Browse/all/abell.html}} \citep{Abell1989}. This catalog was conducted by a manual all-sky search for overdensities of galaxies on photographic plates. The catalog contains 4073 rich galaxy clusters, with at least 30 galaxies in magnitude range between m$_3$ and m$_3$ + 2, where m$_3$ is the magnitude of the third brightest cluster galaxy.
\end{enumerate}

\subsection{High energy surveys for cross-matching analysis}
We also compared our classification of 3PBC sources with those of two hard X-ray catalogs (energies larger than 10\,keV) and one of the latest releases of the Fermi catalog of $\gamma$-ray sources. The former comparison allows us also to obtain more information about the source classification in particular for the Galactic population, while the latter one allows us to look for any trend between the hard X-ray and the $\gamma$-ray emission for the class of blazars. To carry out this task we used the following catalogs.

\begin{enumerate}
\item The 105 month \textit{Swift}-BAT catalog\footnote{\href{https://heasarc.gsfc.nasa.gov/W3Browse/swift/swbat105m.html}{https://heasarc.gsfc.nasa.gov/W3Browse/swift/swbat105m.html}} \citep{Oh2018} is created from data of a uniform hard X-ray all-sky survey in 14-195 keV band. It was developed using the same detector as the 3PBC catalog, but implementing different source algorithms to build X-ray images, data reduction, and source detection. Over 90\,\% of the sky is covered down to a flux limit of 8.40\,$\times$ 10$^{-12}$\,erg\,cm$^{-2}$\,s$^{-1}$ and over 50\,\% of the sky is covered down to a flux limit of 7.24\,$\times$ 10$^{-12}$\,erg\,cm$^{-2}$\,s$^{-1}$. The catalog provides 1632 hard X-ray sources detected above the 4.8\,$\sigma$ level, presenting 422 new detections compared to the previous version of 70-month \textit{Swift}-BAT catalog \citep{Baumgartner2013}. The catalog contains 1132 extragalactic sources, out of which 379 are Seyfert\,I and 448 Seyfert\,II type galaxies, 361 are Galactic sources and 139 are unidentified sources. Objects in the 105 month \textit{Swift}-BAT catalogue are identified together with their optical counterparts by searching the NED and SIMBAD databases and archival X-ray data (e.g., \textit{Swift-XRT}, \textit{Chandra}, \textit{ASCA}, \textit{ROSAT}, \textit{XMM-Newton}, and \textit{NuSTAR}).


\item The \textit{INTEGRAL} IBIS survey hard X-ray catalogue\footnote{\href{https://heasarc.gsfc.nasa.gov/W3Browse/all/ibiscat.html}{https://heasarc.gsfc.nasa.gov/W3Browse/all/ibiscat.html}}, \citep{Bird2016}\footnote{\href{https://heasarc.gsfc.nasa.gov/W3Browse/integral/intibisass.html}{https://heasarc.gsfc.nasa.gov/W3Browse/integral/intibisass.html}} consists of 939 sources detected above a 4.5\,$\sigma$ significance threshold in energy band 17\,--\,100, using the (IBIS) hard X-ray telescope \citep{Winkler2003}. The catalog showed 120 previously undiscovered soft $\gamma$-ray emitters. We also checked our results by comparing them to the findings in \cite{Krivonos2022}.

\item The second release of the fourth \textit{Fermi}-LAT catalog of $\gamma$-ray sources\footnote{\href{https://heasarc.gsfc.nasa.gov/W3Browse/fermi/fermilpsc.html}{https://heasarc.gsfc.nasa.gov/W3Browse/fermi/fermilpsc.html}} (4FGL-DR2, \citep{Ballet2020}, using the Large Area Telescope (LAT) on the \textit{Fermi} Gamma-ray space telescope mission \citep{Atwood2009}, reports 723 new sources, increasing up to 5064 $\gamma$-ray sources. The catalog processed the first 10 years of the data in the energy range between 50 MeV to 1 TeV. The largest class of Galactic sources in the 4FGL-DR2 is constituted by PSRs listing 292 sources, while the extragalactic sample is dominated by blazars with 2226 identified and/or associated BL Lac objects and Flat spectrum radio quasars, and 1517 additional blazar candidates of uncertain type.
\end{enumerate}

\subsection{Multifrequency catalogs for low energy associations}

\begin{enumerate}
\item The current 5$^{th}$ edition of Roma-BZCAT catalog of blazars based on multi-frequency surveys and extensive review of literature\footnote{\href{http://www.ssdc.asi.it/bzcat}{http://www.ssdc.asi.it/bzcat}} \citep{Massaro2015} lists coordinates and multifrequency data for 3561 sources which are either confirmed blazars or sources exhibiting blazar-like behavior. All sources included in the Roma-BZCAT are detected at radio frequencies. According to the Unified AGN model \citep{Antonucci1993, Urry1995}, blazars are AGNs whose jet happens to be closely aligned with our line of sight, exhibiting strong variations, apparent superluminal motion, and emission extending across all electromagnetic spectrum.

\item The Turin-SyCAT \citep{Herazo2022} multifrequency catalog of Seyfert galaxies was built using optical, infrared, and radio selection criteria. Seyfert galaxies are AGNs, which are distinguished as type 1 and type 2 based on the observer's angle \citep{Antonucci1985}. All objects included in its 1$^{st}$ release have an optical spectroscopic classification, allowing us to establish precisely their redshifts and class. The catalog presents 351 Seyfert galaxies, out of which 233 are type 1 and 118 are type 2. In the analysis presented here, the 2$^{nd}$ release of the Turin-SyCAT, increased their number substantially by 80\% to 633 Seyfert galaxies. Details can be find in \hyperref[sec:section5]{Section~5}. All Turin-SyCAT sources with a 3PBC counterpart are detected in the 3PBC at a signal-to-noise ratio above 6.
\end{enumerate}

\section{Classification}
\label{sec:section3}
\begin{figure}
\centering
\includegraphics[width=\hsize]{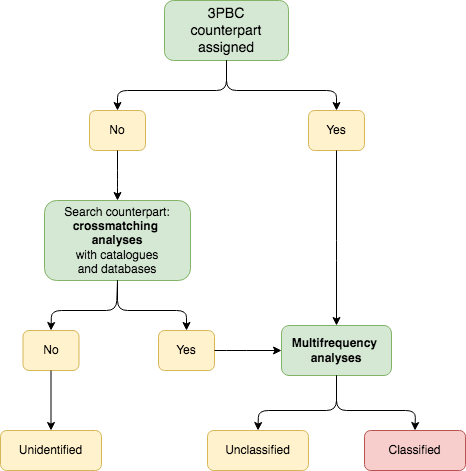}
  \caption{The decision tree adopted in our analysis to distinguish between classified, unclassified, and unidentified hard X-ray sources (see Section 3.1 for a complete description).}
     \label{fig:flowch_scheme}
\end{figure}

For classifying the sources considered in the presented analysis, we adopted the following step-by-step analysis, as shown in Figure~\ref{fig:flowch_scheme} and according to the criteria outlined below. It is worth noting that we are not associating 3PBC sources with their low energy counterparts, but we only update the classification of the associated counterpart based on the latest release of several multifrequency catalogs as those previously listed, and/or follow-up observations that were performed after the 3PBC release \citep[see e.g.,][and references therein]{Molina2009, Malizia2010, Malizia2016, Landi2017, Ricci2017, Koss2017}.

\subsection{Classification scheme}
We start inspecting the 3PBC \citep{Cusumano2010} catalogue. If a 3PBC source has an assigned counterpart we just adopted the multifrequency criteria reported below in this section to classify it. Then, for sources belonging to the extragalactic population, we also verified its redshift estimate. In particular, for all extragalactic sources, being the main focus of the current analysis, the presence of the optical spectrum or a description of it published in the literature is mandatory to consider it as \textit{classified}.

For all sources lacking an assigned low energy counterpart in the original 3PBC, thus being unassociated, we perform the cross-matching analysis with all catalogs reported in section \ref{sec:section2} and we also checked updated information in NED and SIMBAD databases, if any. If no reliable counterpart is found within the BAT positional uncertainty region, we flagged the 3PBC source as \textit{unidentified}. On the other hand, if a potential counterpart is found, as in the previous step, we adopted the multifrequency criteria to classify it and eventually provide a redshift estimate, and, when successful, the associated source is indicated as \textit{classified}.

Moreover, all associated sources that do not have optical spectrum available for their low energy counterpart and/or lack relevant information to determine their classification were labeled as \textit{unclassified}.

\begin{figure}
\centering
\includegraphics[width=\hsize]{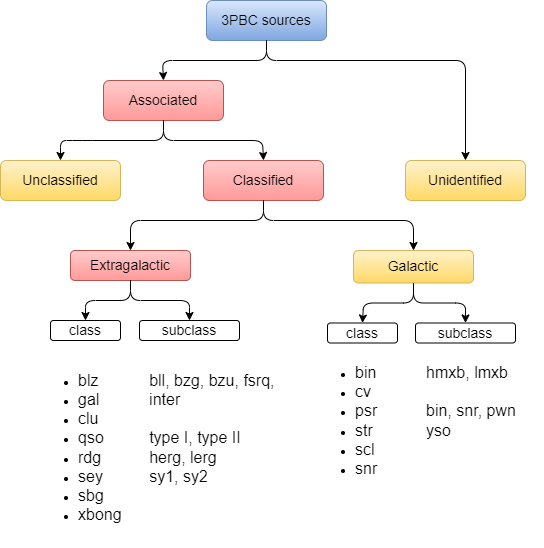}
  \caption{Flowchart of the classification classes. Red cells represent categories leading to extragalactic sources, yellow are categories ending in non-extragalactic sources. Details about the numbers of sources in each class and subclass are described in Table \ref{table:extragalpop} and Table \ref{table:galpop}. Note that classes and subclasses with only a few members are not included for clarity reasons.}
     \label{fig:flowch_classification}
\end{figure}

All classified sources were then split into two main samples distinguishing between Galactic and extragalactic populations. The sky distribution for both the Galactic and the extragalactic populations are shown in Figure~\ref{fig:flowch_classification} and compared in Figure~\ref{fig:piechart_main}. We identified 9 classes and a few sub-classes for both the extragalactic and the Galactic sources discussed in detail in the following subsections, see Table \ref{table:extragalpop} and Table \ref{table:galpop}. 

\subsection{Criteria, classes \& distributions}

\subsubsection{Extragalactic sources}
The largest fraction of sources identified in the extragalactic hard X-ray sky belongs mainly to the two classes of Seyfert galaxies and blazars (\cite{Oh2018}, \cite{Paliya2019}, \cite{Ajello2009}), of which the latter account for 10\% - 20\% of the entire survey population \citep{Diana2022}. Thus, given the possibility to use both the Roma-BZCAT and the Turin-SyCAT \citep{Herazo2022}, built on the basis of multifrequency criteria, for all 3PBC classified sources that belong to these two catalogs we adopted the same classification reported therein. Moreover, we also used their classification schemes to identify new blazars and Seyfert galaxies, to help future releases.

Blazars (class symbol: \textit{blz}) are the largest known population of $\gamma$-ray sources \citep{Abdo2010b, Massaro2012, Abdollahi2020}, dominated by non-thermal radiation over the whole electromagnetic spectrum \citep{Urry1995, Massaro2009}. Their observational features also include high and variable polarization, superluminal motions, very high observed luminosities coupled with a flat radio spectrum \citep{Healey2007, Hovatta2012} peculiar infrared colors \citep{Massaro2011, Abrusco2012, Massaro2013c, Massaro2014} and a rapid variability from the radio to X-ray bands with weak or absent emission lines \citep{Stickel1991}. \cite{Blandford1978} suggested that radiation of blazars could be interpreted as arising from a relativistic jet closely aligned with the line of sight. Blazars were thus classified into 4 categories: BL Lac objects (subclass symbol: \textit{bll}), with featureless optical spectra or presenting only relatively weak and narrow emission lines mainly due to their host galaxies. BL Lacs with their optical-UV spectral energy distribution dominated by the emission of their host galaxy has a subclass symbol: \textit{bzg}. Flat spectrum radio quasars (subclass symbol: \textit{fsrq}) show typical broad emission lines over a blue continuum. Sources exhibiting blazar-like broad-band features, but lacking optical spectroscopic classification are classified as blazars of uncertain type (subclass symbol: \textit{bzu}). According to the nomenclature of the Roma-BZCAT BL Lacs and FSRQs are labeled as BZBs and BZQs, respectively, while, to avoid confusion here they are marked with the classification symbols \textit{bll} and \textit{fsrq}. This choice was adopted because, given the recent optical spectroscopic campaigns devoted to the search for $\gamma$-ray blazars \citep{Landoni2015, Massaro2014, Ricci2015, Herazo2017, Paiano2017, Herazo2019, Paiano2020} a few more blazars, not yet listed in the Roma-BZCAT, were found as low energy counterparts of 3PBC sources and thus, to avoid confusion, we did not use the Roma-BZCAT nomenclature. No further BZGs and/or BZUs were discovered in our analysis and thus no different classification symbols with respect to those of the Roma-BZCAT were used in these cases.

Names for blazar-like counterparts of 3PBC sources were collected from the Roma-BZCAT if the source is listed therein otherwise in the final table the name reported in one of the major radio surveys as NVSS \citep{Condon1998} and/or SUMSS \citep{Bock1999, Mauch2003}, taken from the NED database.

Seyfert galaxies (class symbol: \textit{sey}) were originally defined mainly by their morphology \citep{Seyfert1943} as galaxies with high surface brightness nuclei. Nowadays, they are identified spectroscopically as (mostly spiral) galaxies with strong, highly ionized emission lines. Seyfert galaxies come in two flavors distinguished by the presence (or absence) of broad lines emission in their optical spectra \citep{Khachikian1971, Khachikian1974}. Type 1 Seyfert galaxies (subclass symbol: \textit{sy1}) have both narrow and broad emission lines superimposed to their optical continuum. The former lines originate from a low-density ionized gas with density ranging between $\sim$10$^3$ and 10$^6$ cm$^{-3}$ and line widths corresponding to velocities of several hundred kilometers per second (e.g. \citep{Vaona2012}), while broad lines are located only in permitted transitions, correspondent to electron densities of $\sim$10$^9$\,cm$^{-3}$ and velocities of 10$^4$\,km\,s$^{-1}$ (e.g. \cite{Kollatschny2013}). Type 2 Seyfert galaxies (subclass symbol: \textit{sy2}) show only narrow lines in their optical spectra \citep[e.g.][]{Weedman1977, Miyaji1992, Capetti1999}.

To classify Seyfert galaxies we adopted all the same criteria reported in the Turin-SyCAT \citep{Herazo2022} in terms of (i) presence of the optical spectrum in the literature, (ii) radio, infrared and optical luminosities, (iii) radio morphology. This was chosen because we include the new Seyfert galaxies discovered here in the 2$^{nd}$ release of the Turin-SyCAT as described in the following sections.

Names for Seyfert-like counterparts of 3PBC sources were collected from the 1$^{st}$ edition of the Turin-SyCAT if the source is listed therein otherwise, are reported in the main table with a NED name taken mainly out of one of the following catalogs: 1RXS \citep{Voges1999}, 2MASSS \citep{Skrutskie2006}, 2MASX \citep{Jarrett2000} and BAT105 \citep{Oh2018}. All Seyfert galaxies were then renamed according to the Turin-SyCAT nomenclature.

All extragalactic sources that did not fall into the blazar and Seyfert classes mainly belong to the other two major classes: quasars and radio galaxies.

Quasars (QSOs), (class symbol: \textit{qso}) are AGNs with bolometric luminosities above $\sim10^{40}~\mathrm{erg\,s^{-1}}$. They have broad spectral energy distribution and are emitting from radio up to hard X-ray energies, having variable flux densities almost at all frequencies, mid-IR emission due to the dusty torus, and broad emission lines superimposed to an optical blue continuum \cite{Schmidt1969}. For this extragalactic source class, we also distinguished type 1 and type 2 QSOs on the basis of the presence of broad emission lines in their optical spectra according to the same criteria adopted for the Seyfert galaxies \citep{Khachikian1974}. Then to distinguish a Seyfert galaxy from a QSO we also considered the same thresholds used to create the Turin-SyCAT \citep{Herazo2022}, indicating QSOs as sources with both (i) radio luminosity above $10^{40}$ erg\,s$^{-1}$ and (ii) mid-IR luminosity estimate at 3.4$\mu$ $m$ above $10^{11} L_{\odot}$. Names for the QSOs counterparts of 3PBC sources were collected from is reported in the final table \ref{tab:main_table} with a NED name taken mainly from the following catalogues: 1RXS \citep{Voges1999}, 2MASSS \citep{Skrutskie2006}, 2MASX \citep{Jarrett2000} and 1SXPS \citep{Evans2014}.

Radio galaxies (RDGs), (class symbol: \textit{rdg}) are radio-loud AGNs whose radio emission is at least 100 times that of normal elliptical galaxies and extends beyond tens of kpc scale\citep{Urry1995, Moffet1966, Massaro2011, Velzen2012}, thus being neatly distinct from the Seyfert galaxies. On the other hand, to distinguish between QSO and RDG we adopted a radio morphological criterion where the latter clearly presents diffuse radio emission at a large scale when radio maps are available to check it. We used the same criteria and classification scheme recently adopted by \citep{Capetti2017a, Capetti2017b}. If the source was not listed with those names, we took the NED name mainly from 3C \citep{Spinrad1985}, 4C \citep{Pilkington1965, Gower1967} or 7C \citep{Hales2007} catalogs.

We firstly classified RDGs on the basis of their radio morphologies at 1.4 GHz distinguishing between classical FR\,I and FR\,II sources \citep{Fanaroff1974}. On the other hand, we also considered the two subclasses of radio galaxies defined on the basis of their optical emission lines, distinguishing between high excitation radio galaxies (HERGs), (subclass symbol: \textit{herg}) and low excitation radio galaxies (LERGs), (subclass symbol: \textit{lerg}) \citep{Hine1979}. HERGs are almost always FRIIs, while LERGs can be either FRIs or FRIIs \citep{Buttiglione2010}.

We also considered galaxy clusters (class symbol: \textit{clu}) as extragalactic sources of hard X-rays. Galaxy clusters are the largest gravity-bounded structures in the Universe, composed primarily of dark matter, highly ionized and extremely hot intra-cluster gas of low density, and galaxies \citep{Sarazin1986, Giodini2009}. Their X-ray emission is mainly due to bremsstrahlung radiation of relatively hot particles in their intra-cluster medium in the soft X-rays \citep[i.e., between 0.5 and 10 keV][]{Nevalainen2003}, although a tail of this emission is also detectable at higher energies \citep{Ajello2010}. Since it is well known that some galaxy clusters were also detected by the BAT instrument on board SWIFT \citep{Ajello2010} we reported 3PBC sources associated with them mainly when the cross-match with the Abell catalog indicated the possible presence of a galaxy cluster within the hard X-ray positional uncertainty region.

Finally, we highlight that a handful of extragalactic sources, not belonging to the five major classes listed above, fall into the following categories, being classified as starburst galaxies (class symbol: \textit{sbg}), \citep{Searle1973, Weedman1981}, galaxies forming stars at unusually fast rates (10$^3$ times faster than in an average galaxy), X-ray bright optically normal galaxies (class symbol: \textit{xbong}), which are normal galaxies, not hosting an AGN, but having substantial X-ray luminosity \citep{Elvis1981, Comastri2002, Yuan2004}, low-ionization nuclear emission-line region galaxies (class symbol: \textit{liner}) \citep{Singh2013} and normal galaxies (class symbol: \textit{gal}), the latter not hosting an AGN but in a few cases interacting with nearby companions. Names of the 3PBC counterparts for those sources were collected mainly from 2MASX \citep{Jarrett2000} and 2MASS \citep{Skrutskie2006} catalogues.

\begin{figure*}[ht]
\begin{center}
\includegraphics[height=5.8cm,width=8.cm]{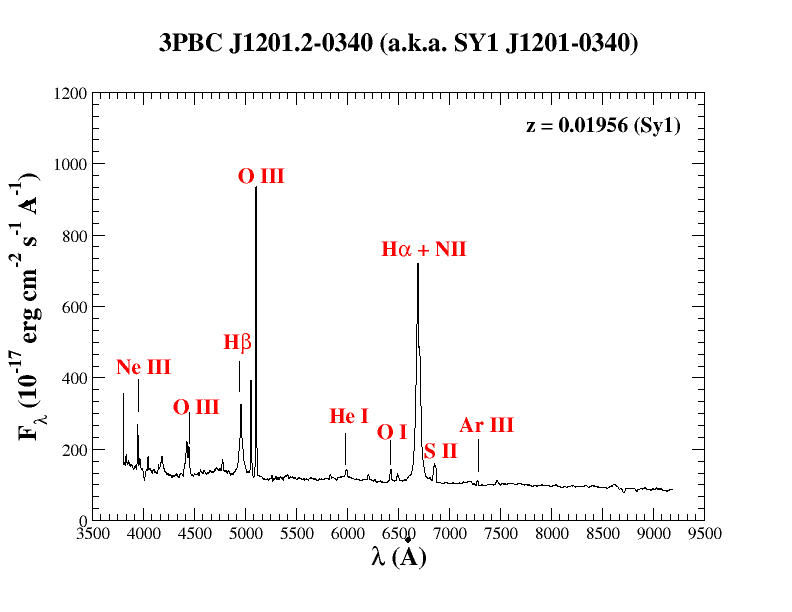}
\includegraphics[height=5.8cm,width=8.cm]{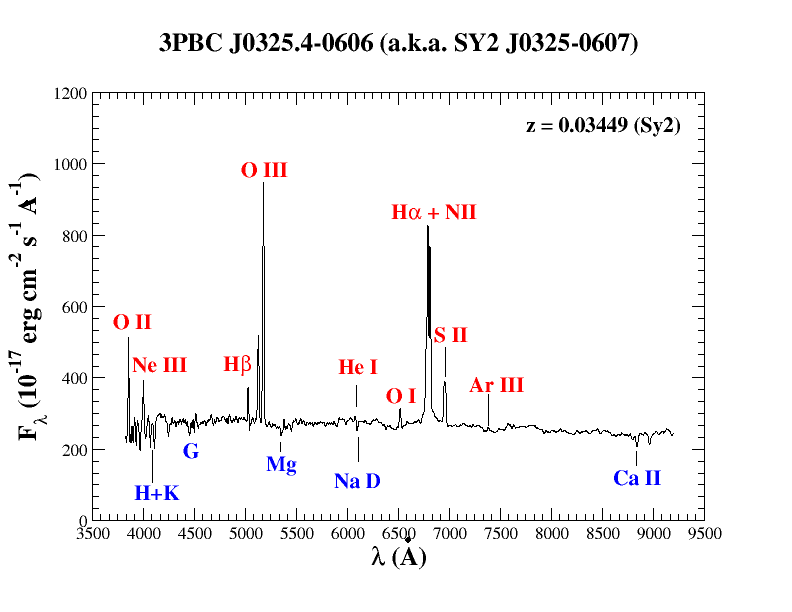}

\includegraphics[height=5.8cm,width=8.cm]{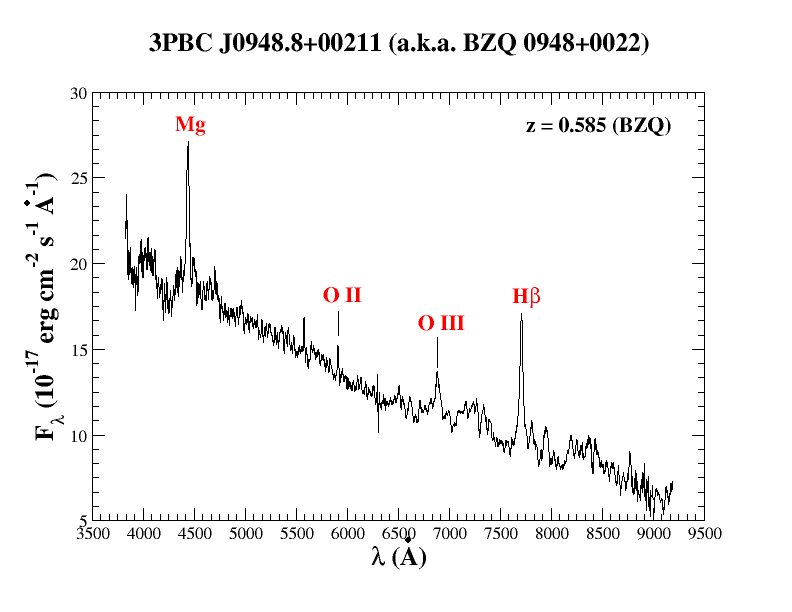}
\includegraphics[height=5.8cm,width=8.cm]{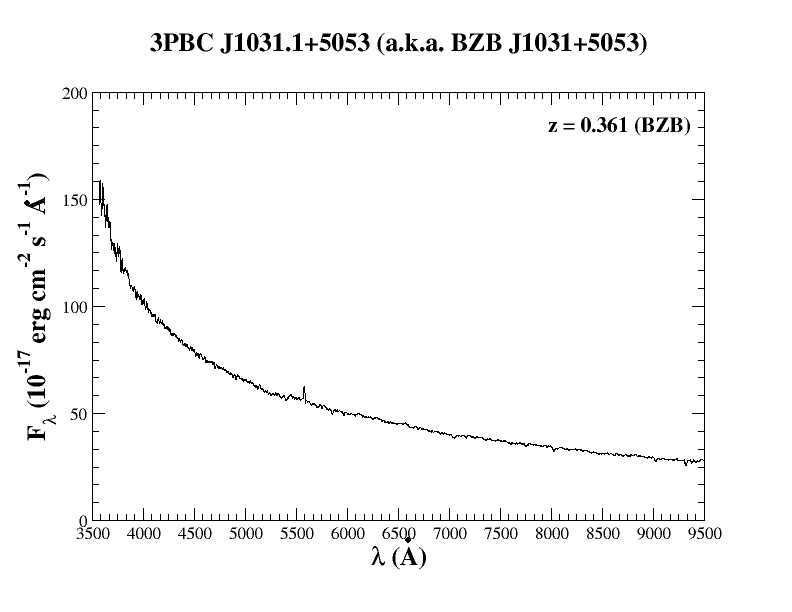}

\end{center}
\caption{Four images showing examples of the optical spectra of 3PBC counterparts identified in our refined analysis.
Top left: Type I Seyfert galaxy 3PBCJ1201.2-0340. 
Top right: Type II Seyfert galaxy 3PBCJ0325.4-0606. 
Middle left: Flat spectrum radio quasar (fsrq) 3PBCJ0948.8+0021. 
Middle right: BL Lac object 3PBCJ1031.1+5053 (BZB in Roma-BZCAT \citep{Massaro2009} nomenclature).
}
\label{fig:spectra_examples}
\end{figure*}

\begin{figure*}[ht]
\begin{center}
\includegraphics[height=5.8cm,width=8.cm]{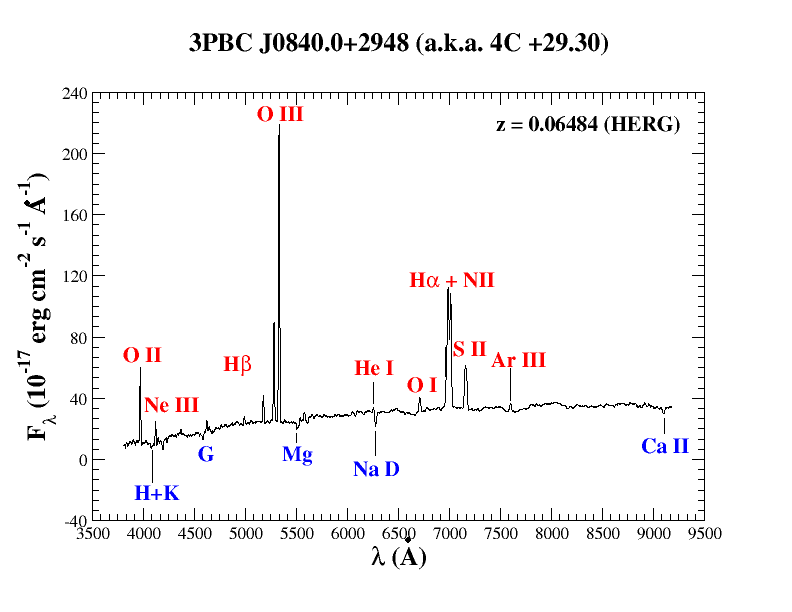}
\includegraphics[height=5.8cm,width=8.cm]{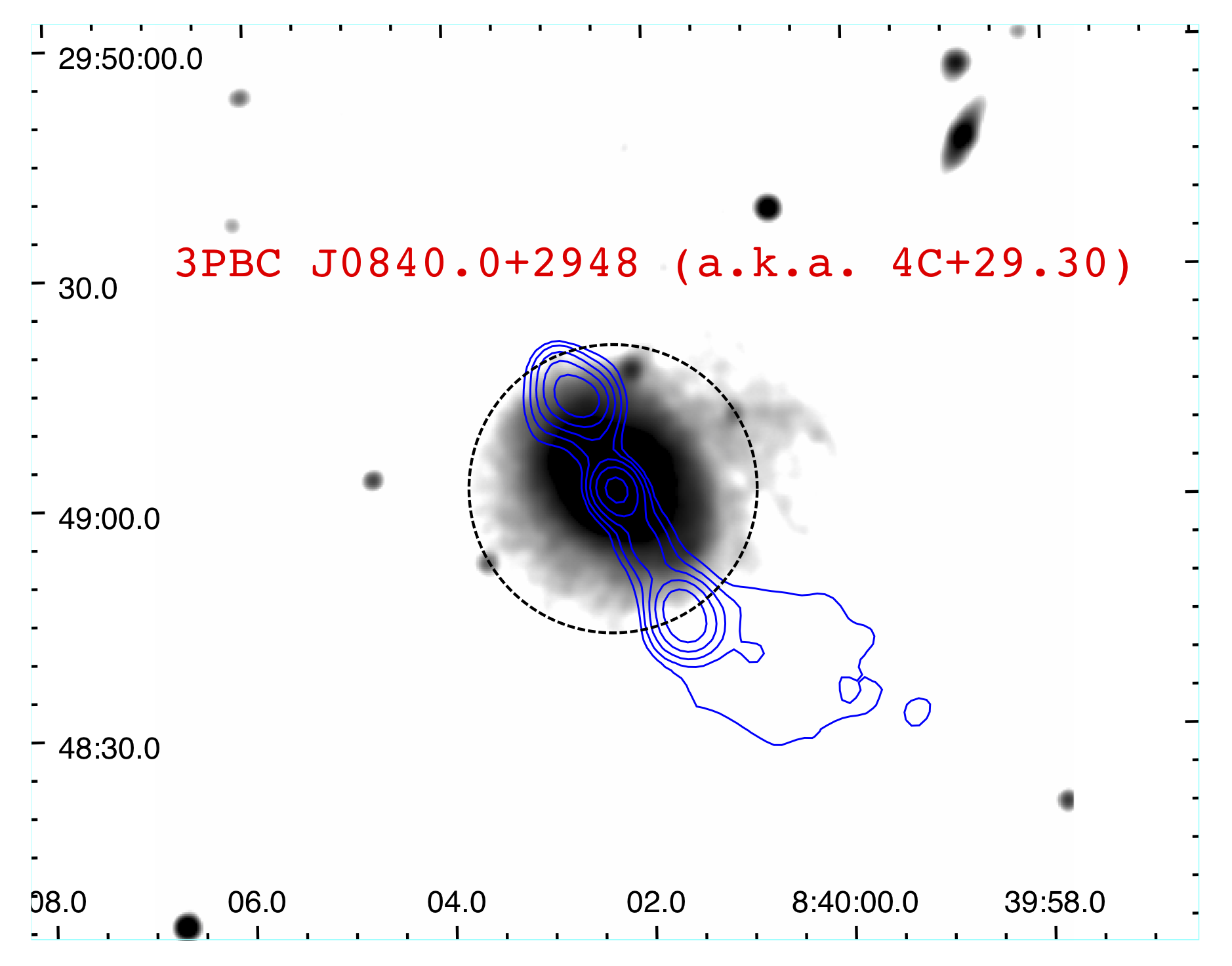}
\end{center}
\caption[caption for spectra_examples_2]{Left panel: the optical spectrum of the HERG 4C\,+29.30 associated with 3PBC\,J0840.0+2948 where main spectral emission and/or absorption lines are marked. Right panel: the optical image from the Pan-STARRS archive in the $r$ band of 4C\,+29.30 with radio contours overlaid, drawn from the 3 GHz VLASS\footnote{https://science.nrao.edu/science/surveys/vlass} radio map extending beyond the host galaxy.}
\label{fig:spectra_examples_2}
\end{figure*}

We list a preview of the first 10 sources included in Table\ref{tab:main_table}, our revised version of the 3PBC catalog in which we provide the 3PBC catalog name, coordinates, counterpart name, counterpart coordinates, spectroscopic redshifts, the classification in our class and subclass system and the WISE counterpart name. We show examples of spectra of a few objects in Figure\ref{fig:spectra_examples} and Figure\ref{fig:spectra_examples_2}.

\begin{table}[h]
\caption{Numbers of 3PBC extragalactic sources associated in each class and subclass.} 
\label{table:extragalpop}      
\begin{tabular}{cccc}
 \hline 
 \hline
  Class    & Class   & Subclass &  Subclass \\
  symbol  & number   & symbol  &  number   \\
\hline
blz & 129 & bll & 30 \\
&  & bzg & 7 \\
&  & bzu & 24 \\
&  & fsrq & 68 \\
gal & 10 & interacting & 3 \\
&  & - & 7 \\
clu & 27 & & \\
liner & 1 & & \\
qso & 26 & type 1 & 18 \\
&  & type 2 & 1 \\
&  & ? & 7 \\
rdg & 25 & herg & 21 \\
&  & lerg & 3 \\
&  & ? & 1 \\
sey & 593 & sy1 & 325 \\
&  & sy2 & 268 \\
sbg & 5 & & \\
xbong & 4 & & \\
\hline
\end{tabular}

Note: This is the classification of the 3PBC Galactic sources according to our classification scheme.

\end{table}

\subsubsection{Galactic sources}
In our Milky Way, most of the sources emitting in the hard X-rays are X-ray binaries \citep{Grimm2002}, while the second dominant class of hard X-ray sources is the cataclysmic variables \citep{Revnivtsev2008}.

X-ray binaries (BINs) (class symbol: \textit{bin}) are systems of double stars containing compact stellar remnants, such as neutron stars, pulsars, or black holes, and a normal star which can range a variety of masses (e.g. \cite{Charles2003, Knigge2011}). The compact stellar remnant accretes material from its companion, creating continual or transient X-ray emissions. X-ray binaries are classified based on their companion star distinguishing between low mass X-ray binaries (subclass symbol: \textit{lmxb}) having the companion star of mass $\lesssim$\,1\, M$_{\odot}$ and high mass X-ray binaries (subclass symbol: \textit{hmxb}) usually accompanied by a star of mass $\gtrsim$\,10\, M$_{\odot}$, where the accretion happens directly from a stellar wind of the companion star. Names for the BINs counterparts of 3PBC sources were collected mainly from the following catalogues: IGR \citep{Bird2004}, 1H, SWIFT \citep{Ajello2010} and (RX+XTE+SAX) \citep{Bade1992, Voges1999, Frontera2009}.

Cataclysmic variables (CVs), (class symbol: \textit{cv}) are binary systems composed of a main-sequence companion star and a compact stellar remnant which is a white dwarf (WD) \citep{Revnivtsev2008}. The accretion happens almost always via filling the Roche-lobe of the companion star and subsequent formation of an accretion disk around the WD \citep{Warner1995}. Their X-ray emission can originate from a variety of processes depending on the type of the CV. CVs which do not have strong magnetic fields accrete matter closer to the surface of the WD and produce sporadic eruptions. 
For 4 sources belonging to the CV class, we also indicated if they are symbiotic stars or novae, however, given their relatively low number with respect to all CVs identified we did not label these as subclasses and we only report the source class. Names for the CVs counterparts of 3PBC sources were collected mainly from the following catalogues: CV \citep{Downes2005}, IGR \citep{Bird2004}, 1RXS \citep{Voges1999} and 2MASS \citep{Skrutskie2006}.

The hard X-ray sky is also populated by isolated X-ray pulsars (PSR), (class symbol: \textit{psr}), not being hosted in X-ray binaries. Since they can be indeed hosted in pulsar wind nebulae (subclass symbol: \textit{pwne}) or supernova remnants we highlight the presence of this extended emission around the PSR in the subclass column. On the other hand, if the hard X-ray emission is indeed due to a supernova remnant not hosting a neutron star then we adopted a different class (subclass symbol: \textit{snr}), in these cases, their hard X-ray emission is due to the thermal radiation of plasma heated in shocks, coupled with non-thermal synchrotron radiation (see e.g., \cite{Vink2012}). Names for the PSRs counterparts of 3PBC sources were collected mainly from the ATNF PSR catalog or other radio surveys.

As occurred for the extragalactic hard X-ray population a handful of sources were also identified belonging to normal stars (class symbol: \textit{str}), (coming with a subclass: \textit{yso} for young stellar objects) and star clusters (class symbol: \textit{scl}). X-ray emission from main sequence stars of masses >\,10\, M$_{\odot}$ can be due to discrete ionized metal lines in their spectrum. Young stellar objects, protostars, and T Tauri stars also exhibit X-ray radiation, predominantly emerging from magnetic coronae accreting material where shocks occur \citep{Gudel2009}. 
On the other hand, star clusters can appear as an amalgamation of point-like sources and extended X-ray emissions. Their point-like component can be produced by hot stars and/or SNR, lasting a few thousand years while their extended component is produced by star cluster wind, formed by the interaction of stellar winds of massive O or B type stars, Wolf-Rayet stars, and supernovae explosions (e.g. \cite{Cant2000, Law2004, Oskinova2005}). In addition to them we also reported the classification for one microquasar (class symbol: \textit{mqso}), namely: 3PBC J0804.7-2748. Microquasars are similar to quasars but in a much smaller case. Their radiation comes from a stellar mass black hole or a neutron star accreting matter from a normal star \citep{Mirabel2010}. In addition, we report one planetary nebula (class symbol: \textit{pn}): 3PBC J1701.5-4306. Planetary nebulae are the ejected red giant's atmosphere ionized by the leftover star's core, forming at the end of life of stars with initial masses in the range $\sim$ 1 to 8 solar masses. Lastly, we also labeled the Galactic center Sgr A$^*$ with the symbol: \textit{galcent}.

\begin{table}[h]
\caption{Numbers of the Galactic 3PBC sources associated in each class and subclass.}
\label{table:galpop}      
\begin{tabular}{cccc}
 \hline \hline
  Class    & Class        & Subclass &  Subclass \\
  symbol  & number   & symbol  &  number   \\
\hline
bin & 231 & hmxb & 117 \\
&  & lmxb & 108 \\
&  & ? & 6 \\
cv & 83 & & \\
str & 12 & - & 6 \\
&  & yso & 1 \\
&  & ? & 1 \\
psr & 21 & bin & 1 \\
&  & - & 5 \\
&  & snr & 12 \\
&  & pwn & 3 \\
scl & 2 & & \\
snr & 4 & & \\
mqso & 1 & & \\
pn & 1 & & \\
galcent & 1 & & \\
\hline

\end{tabular}

Note: This is the classification of the 3PBC Galactic sources according to our classification scheme. Classes: galcent, mqso, and pn were omitted due to a small member count (1 each).

\end{table}

\subsubsection{Sky distributions}

\begin{figure}[b]
\centering
\includegraphics[width=\hsize]{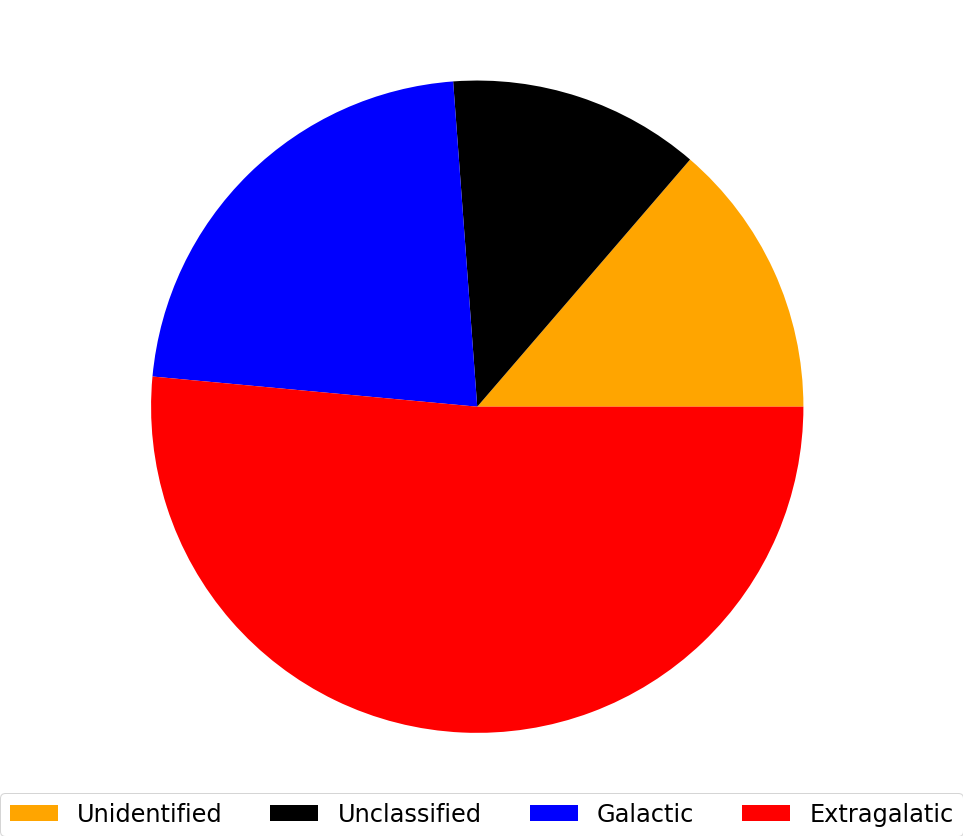}
  \caption{Pie-chart showing the fractions of the main classification categories derived thanks to our revised analysis of the 3PBC. The representation of the categories is following: 218 Unidentified ($\sim$\,13.7\%), 199 Unclassified ($\sim$\,12.5\%), 356 Galactic ($\sim$\,22.3\%), 820 Extragalactic ($\sim$\,51.5\%).}
     \label{fig:piechart_main}
\end{figure}

Starting from the total number of 1593 sources listed in the 3PBC catalogue \citep{Cusumano2010} we found that according to our analysis there are 218 unidentified hard X-ray sources ($\sim$13.7\,\%) and 199 unclassified sources ( $\sim$12.5\,\%), see Figure~\ref{fig:piechart_main}. The classified sources are distinguished into two main groups: Galactic objects including 356 sources ($\sim$22.2\,\%), and extragalactic objects having 820 sources ($\sim$51.5\,\%).

\begin{figure*}
\centering
\includegraphics[width=\textwidth]{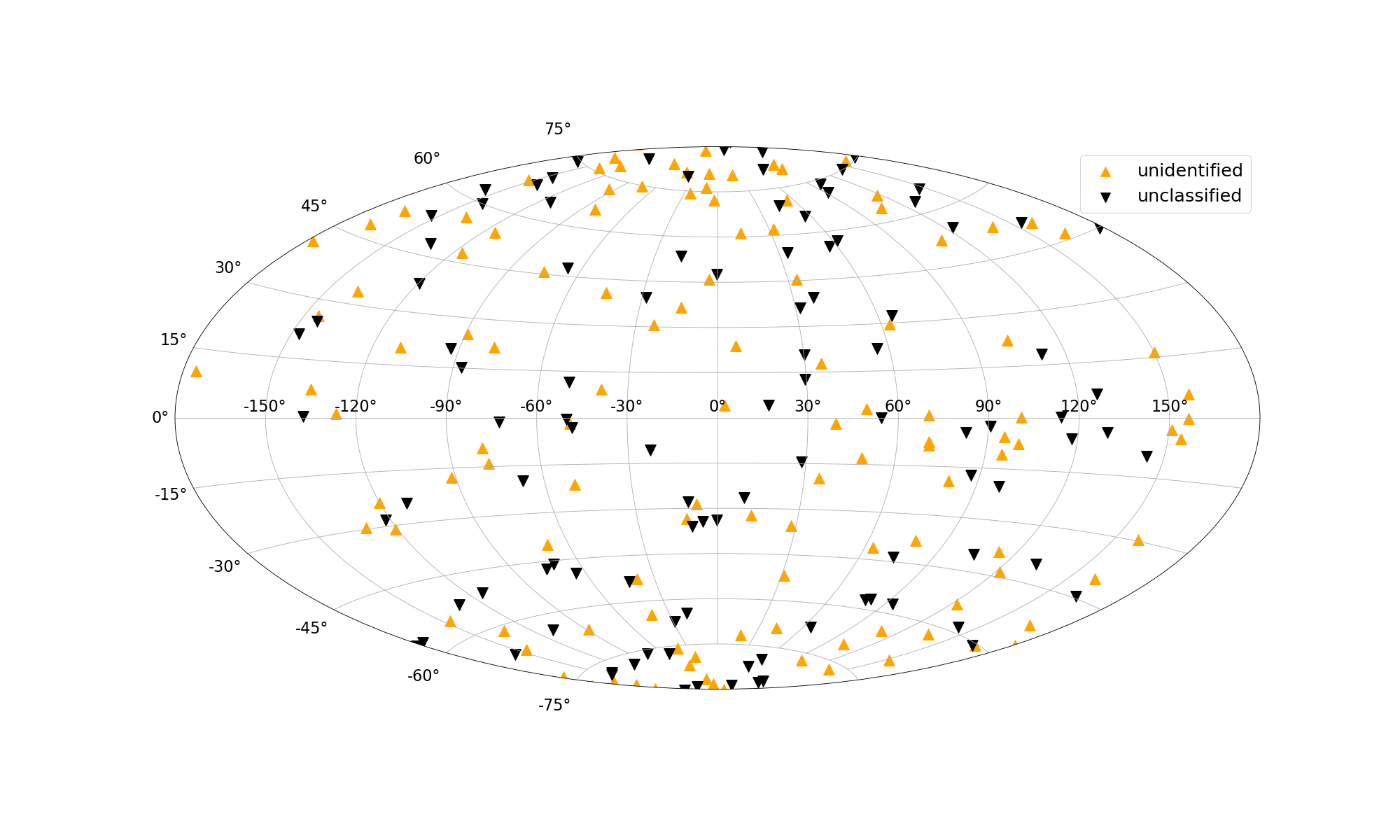}
  \caption{Hammer-Aitoff projection derived thanks to our revised analysis of the 3PBC, showing unidentified and unclassified sources.}
     \label{fig:hammer_aitoff_main_table_uni_unc}
\end{figure*}

\begin{figure*}
\centering
\includegraphics[width=\textwidth]{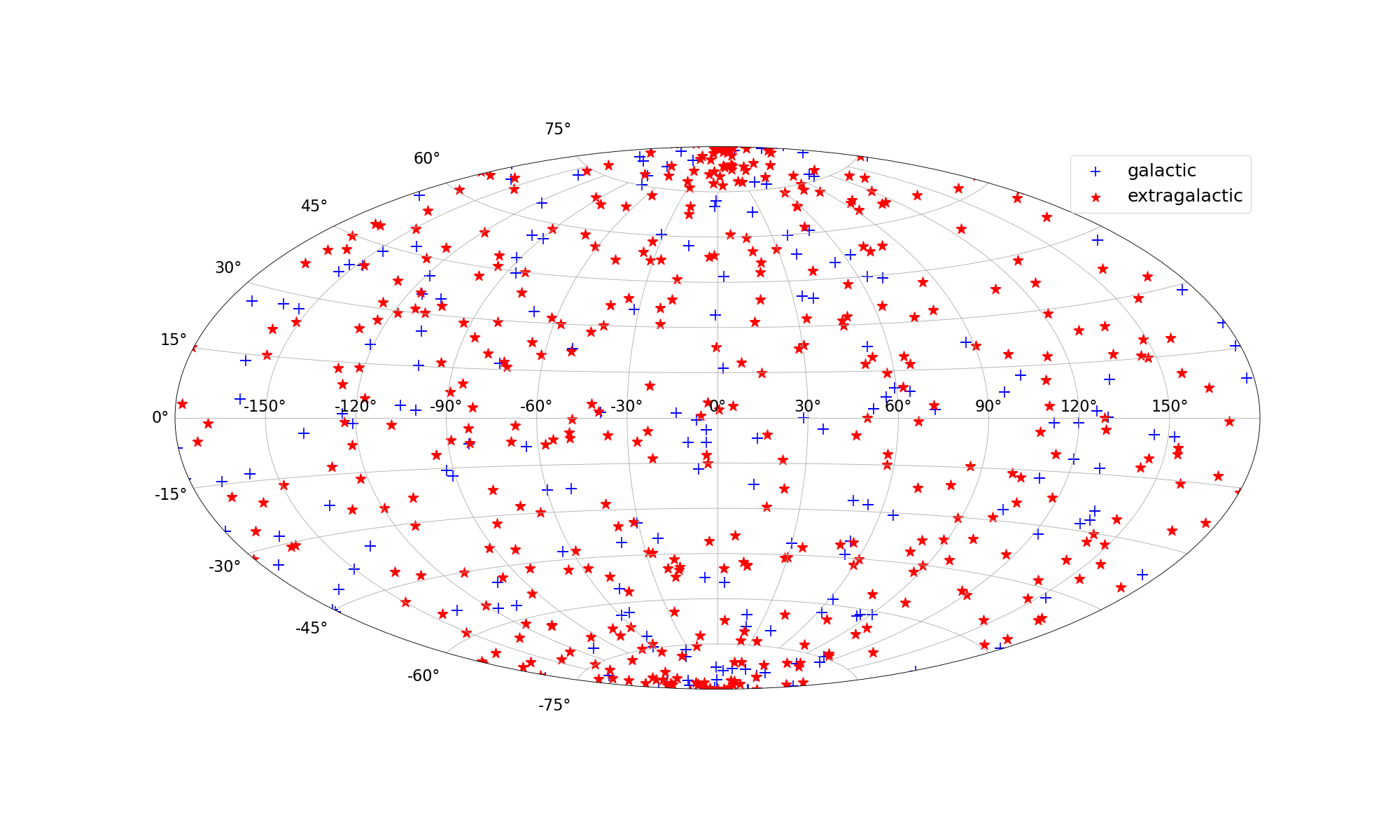}
  \caption{Hammer-Aitoff projection based on our revised analysis of the 3PBC, showing Galactic and extragalactic sources.}
     \label{fig:hammer_aitoff_main_table_gal_extgal}
\end{figure*}

\begin{figure}
\centering
\includegraphics[width=\hsize]{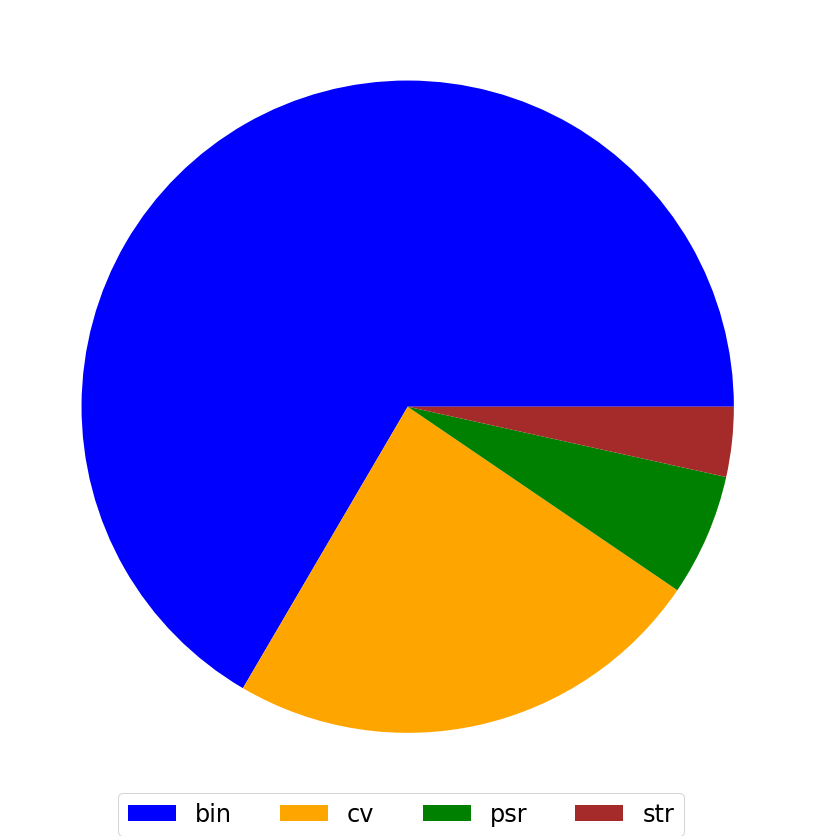}
  \caption{Pie-chart showing the fractions of galactic classes derived thanks to our revised analysis of the 3PBC. The classes are represented as follows: 231 bin ($\sim$\,66.6\%), 83 cv ($\sim$\,23.9\%), 21 psr ($\sim$\,6.1\%), 12 str ($\sim$\,3.5\%). Note, that classes galcent, mqso and pn are omitted for a small contribution (1 member each).}
     \label{fig:piechart_gal}
\end{figure}

\begin{figure}
\centering
\includegraphics[width=\hsize]{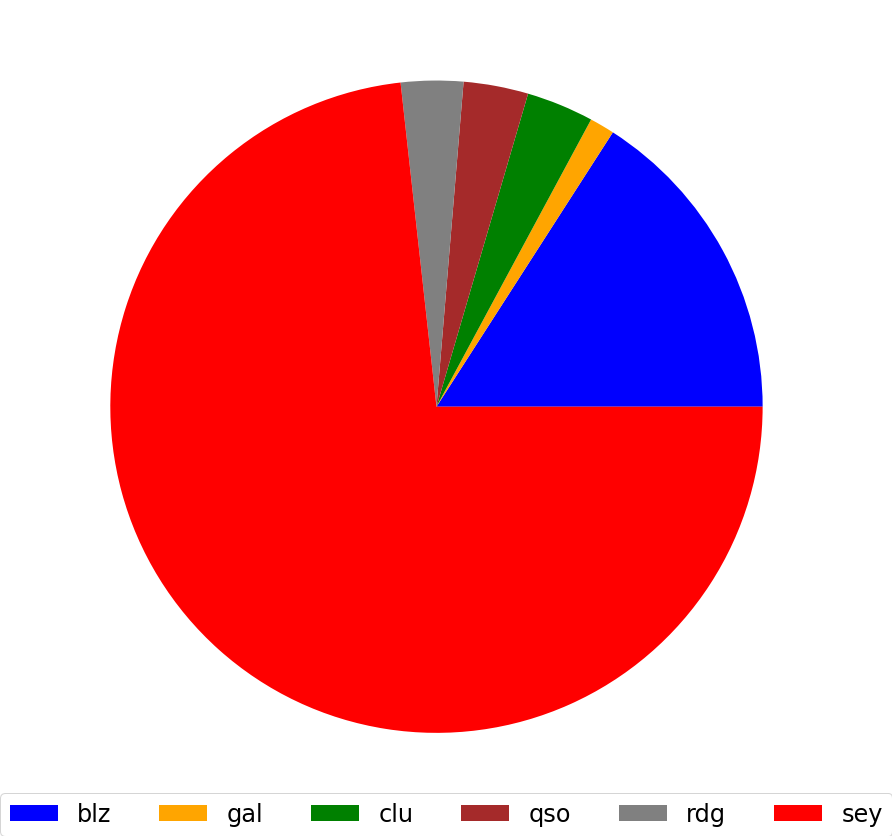}
  \caption{Pie-chart showing the fractions of extragalactic classes derived thanks to our revised analysis of the 3PBC. Individual classes have following representation: 129 blz ($\sim$\,15.9\%), gal 10 ($\sim$\,1.2\%), 27 clu ($\sim$\,3.3\%), 26 qso ($\sim$\,3.2\%), 25 rdg ($\sim$\,3.1\%), 593 sey ($\sim$\,73.2\%). Note that the liner, xbong, and sbg subclasses are omitted for a small contribution (1, 4, and 5 members each, respectively).}
 \label{fig:piechart_extragal}
\end{figure}

We show the sky distribution of 3PBC sources via the Hammer-Aitoff projection for both unclassified and unidentified cases in Figure~\ref{fig:hammer_aitoff_main_table_uni_unc} and for classified sources, distinguishing between Galactic and extragalactic ones in Figure~\ref{fig:hammer_aitoff_main_table_gal_extgal}.
Given the source distributions for both unidentified and unclassified sources that appear to be quite uniform over the whole sky, we could expect that a large fraction of them could have an extragalactic origin. This could imply that the lack of classified counterparts is mainly due to missing follow-up spectroscopic observations, thus strengthening the need to complete optical campaigns carried out to date \citep[see e.g.][]{Masetti2006a, Masetti2006b, Masetti2009, Cowperthwaite2013}.
Then fractions of other classes for extragalactic sources are shown in Figure~\ref{fig:piechart_extragal} and for Galactic classes in Figure~\ref{fig:piechart_gal}.

\begin{sidewaystable}
    
    \caption{Our revised version of the 3PBC catalog. The entire catalogue table is available in the online material.}
    \label{tab:main_table}
    \begin{tabular}{cccccccccc}
    \hline \hline
    
    3PBC & R.A.$^{3PBC}$ & Dec.$^{3PBC}$ & counterpart & R.A.$^{ctp}$ & Dec.$^{ctp}$ & $z$ & class & subclass & WISE \\
    name &  (deg)        &  (deg)        &  name       & (deg)        & (deg)        &     &       &          & name \\
    
    \hline \hline
    J0000.9-0708 & 0.228  & -7.134  & 2MASS J00004877-0709115  & 0.203216 & -7.153221  & 0.03748  & sey & sy2  & J000048.77-070911.6 \\
    J0001.7-7659 & 0.429  & -76.986 & 2MASX J00014596-7657144  & 0.441917 & -76.953972 & 0.05839  & sey & sy1  & J000146.08-765714.2 \\
    J0002.5+0322 & 0.636  & 3.367   & SY1 J0002+0322           & 0.610046 & 3.351961   & 0.025518 & sey & sy1  & J000226.42+032106.8 \\
    J0002.5+0322 & 0.853  & 27.638  & 2MASX J00032742+2739173  & 0.864283 & 27.654828  & 0.03969  & sey & sy2  & J000327.41+273917.0 \\
    J0002.5+0322 & 1.009  & 70.312  & SY2 J0004+7020           & 1.008228 & 70.32175   & 0.096    & sey & sy2  & J000401.97+701918.3 \\
    J0006.3+2012 & 1.584  & 20.205  & SY1 J0006+2013           & 1.581389 & 20.202968  & 0.025785 & sey & sy1  & J000619.53+201210.6 \\
    J0010.4+1058 & 2.624  & 10.976  & 5BZQ J0010+1058          & 2.629166 & 10.974888  & 0.089100 & blz & fsrq & J001031.00+105829.5 \\
    J0016.7-2611 & 4.194  & -26.2   &                          & 0.       & 0.         & 0.    
    & uhx &      &                     \\
    J0017.4+0519 & 4.37   & 5.326   & HS 0014+0504             & 4.344167 & 5.352778   & 0.11     & sey & sy1  & J001722.71+052111.4 \\
    J0017.8+8135 & 4.454  & 81.591  & 5BZQ J0017+8135          & 4.28525  & 81.58561   & 3.387000 & blz & fsrq & J001708.50+813508.1 \\
    \end{tabular}
    
Only the first 10 lines are reported here. Col. (1) 3PBC source name; Cols. (2,3) Right Ascension and Declination of the 3PBC source (Equinox J2000); Col. (4) name of the counterpart assigned in our refined analysis; col. (5,6) Right Ascension and Declination of the counterpart (Equinox J2000); Col. (7) counterpart redshift if extragalactic; Cols. (8,9) class and subclass assigned according to our classification scheme; Col. (10) WISE name of the counterpart.
\end{sidewaystable}

\section{Characterizing the Extragalactic Hard X-Ray Sky}
\label{sec:section4}
Our revised analysis of the 3PBC lists 820 extragalactic sources, classified into 9 classes: 129 blazars (\textit{blz}), 10 galaxies (\textit{gal}), 27 galaxy clusters (\textit{clu}), 1 low-ionization nuclear emission-line region galaxy (\textit{liner}), 26 quasars (\textit{qso}), 25 radio galaxies (\textit{rdg}), 593 Seyfert galaxies (\textit{sey}), 5 star-burst galaxies (\textit{sbg}) and 1 X-ray bright optically normal galaxy (\textit{xbong}). Table~\ref{table:extragalpop} reports those numbers together with the number of sources in their associated subclasses.

\begin{figure}
\centering
\includegraphics[width=\hsize]{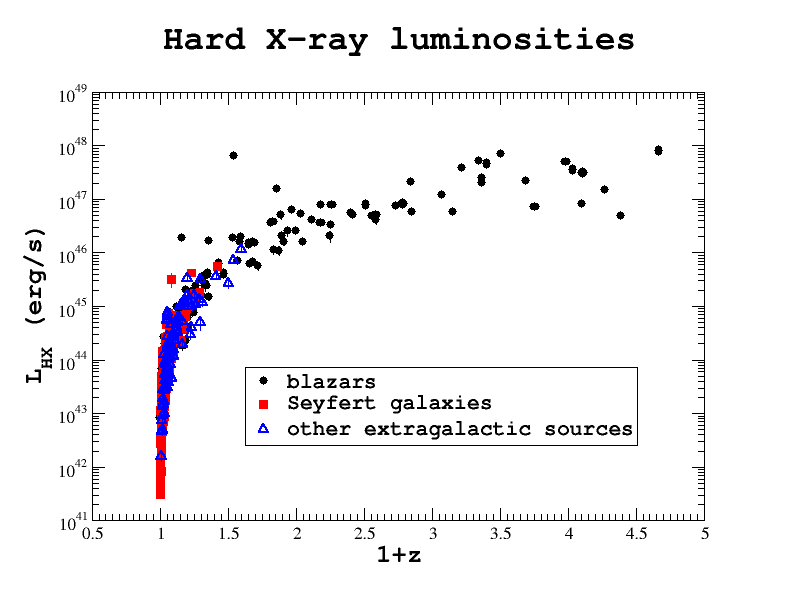}
  \caption{K-corrected hard X-ray luminosity computed using the 15-150 keV flux reported in the 3PBC. Given the high numbers of sources used to compute the correlation coefficients, the p-chance for all correlations is below 10$^{-5}$ level of significance. Seyfert galaxies are reported as red squares, blazars as blue circles while all other extragalactic sources are marked as green triangles. Statistical uncertainties are smaller than the size of the points.}
     \label{fig:luminosity}
\end{figure}

The most abundant class of extragalactic sources are Seyfert galaxies (Figure~\ref{fig:piechart_extragal}), while the second largest population of extragalactic sources emitting in the hard X-rays is constituted by blazars. The hard X-ray luminosity, K-corrected, is shown in Figure~\ref{fig:luminosity} as a function of the redshift with particular emphasis on the two classes of Seyfert galaxies and blazars. We used the measured spectral index reported in the 3PBC for K-correlation computation.

Once we assigned the coordinates of each counterpart we also crossmatched the 3PBC catalog with the AllWISE survey\footnote{\href{https://wise2.ipac.caltech.edu/docs/release/allwise/}{https://wise2.ipac.caltech.edu/docs/release/allwise/}} \citep{Cutri2014} and we found that adopting an association radius of 3.3$\arcsec$, as typically used in other analyses \citep{D'Abrusco2019, Massaro2012b, Menezes2020} we found 1279 mid-IR counterparts in the 1593 3PBC sources. It is worth noting that associating sources within this angular separation corresponds to a chance probability of having a spurious match lower than $\sim$2\,\% \citep{Massaro2013b, Massaro2015c}.

\begin{figure}
\centering
\includegraphics[width=\hsize]{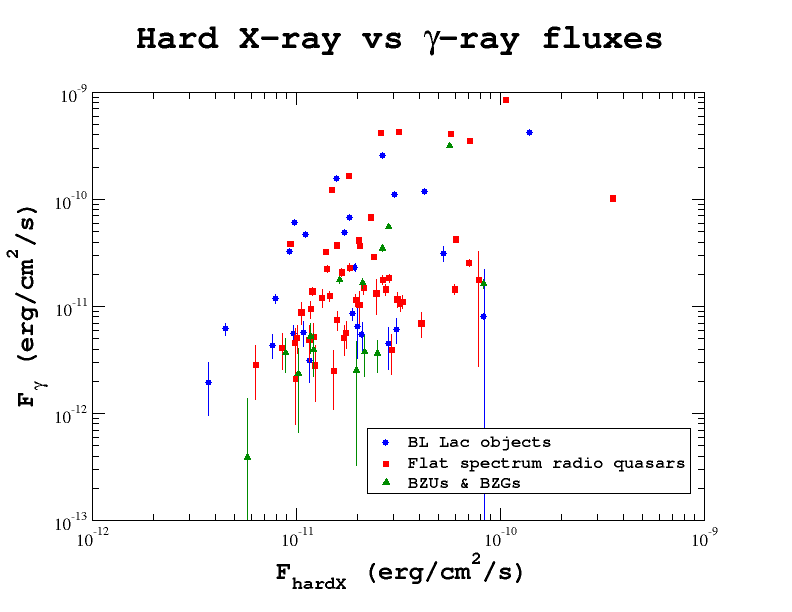}
\includegraphics[width=\hsize]{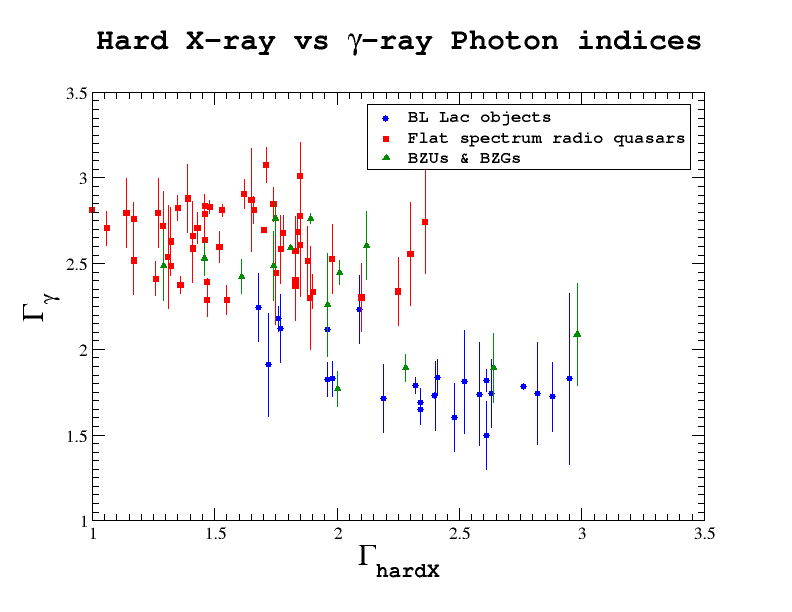}
  \caption{Top: Hard X-ray vs $\gamma$-ray and vs photon indices (bottom) correlation of all blazars in our sample. We can see, that BL Lacs are steeper in hard X-rays and flatter in $\gamma$-rays (bottom), however, their Hard X-ray vs $\gamma$-ray flux distribution (top) appears similar.}
     \label{fig:gamma}
\end{figure}

We also used the counterpart coordinates to carry out a crossmatch between all blazars listed in the 3PBC and those associated with the 4FGL catalog. There are 92 out of 129 blazars with a {\it Fermi} counterpart and with known redshift, with 25 of them belonging to the BL Lac class and 52 to that of FSRQs. For all these $\gamma$-ray emitting blazars we also found two neat trends/correlations between their hard X-ray and $\gamma$-ray emissions as highlighted in Figure~\ref{fig:gamma}. The first trend is between their hard X-ray and $\gamma$-ray fluxes, where a mild correlation is also reported: 0.52 is the measured value for the correlation coefficient for the whole blazar sample. The p-chance for all correlations is below 10$^{-5}$ level of significance due to the high number of sources used to compute the correlation coefficients. Then a second trend was indeed found between the photon indices of blazars measured in the 3PBC and in the 4FGL catalogs.

Both trends highlighted for the blazar population emitting in the hard X-rays are expected given the nature of their emission  (e.g. \cite{Acharyya2021}). For BL Lac objects the steep hard X-ray spectra could be due to emission arising from the tail of their synchrotron \citep {Maraschi1992} component, and the flat $\gamma$-ray spectra are related to the peak of their inverse Compton bump at $\gamma$-ray energies \citep{Maraschi1999, Marscher1985, Dermer1995}. On the other hand for the FSRQs both the hard X-ray and the $\gamma$-ray emission are due to their inverse Compton component peaking in the $\gamma$-ray band \citep{Acharyya2021}. Then we also note that even if the broadband spectral energy distributions (SEDs) of BL Lacs are mainly interpreted as due to Synchrotron Self Compton emission while that of FSQRs to external Compton radiation \citep{Abdo2010} relativistic particles responsible for both SED bumps are the same and thus we could expect that fluxes in the hard X-rays and in the $\gamma$-rays are, on average, connected \citep[see e.g.]{Wolter2008}.

\section{Second Release of the Turin-SyCAT}
\label{sec:section5}
We found 282 new Seyfert galaxies resulting from our analysis of the extragalactic hard X-ray sky presented in the previous sections.
Adding all new Seyfert galaxies to those already included in the $1^{st}$ release of the Turin-SyCAT its $2^{nd}$ release lists 633 Seyfert galaxies: 351 types 1 and 282 types 2. Thus we added a total of 118 types 1 and 164 types 2 Seyfert galaxies and we also present here an updated analysis of the infrared - hard X-ray connection including all new sources.

Sources added in the $2^{nd}$ release of the Turin-SyCAT were selected according to the same procedure as in \citet{Herazo2022}. These strict selection criteria allow us to have a negligible fraction of contaminants since we selected only extragalactic sources with a Seyfert-like optical spectrum and having:
\begin{enumerate}
\item a published optical spectrum;\\
\item a luminosity in Radio lower than $<$10$^{40}$ erg~s$^{-1}$ if a counterpart is listed in the two major radio surveys \cite[i.e., NVSS and SUMSS][respectively]{Condon1998, Mauch2003};
\item a counterpart in the AllWISE Source catalog with a mid-IR luminosity at $3.4 \mu\,m$ less than $ 3 \times 10^{11}$ $L_\sun$. This was mainly adopted to avoid the selection of QSOs.
\end{enumerate}

\begin{figure}[ht]
\begin{center}
\includegraphics[width=8.cm]{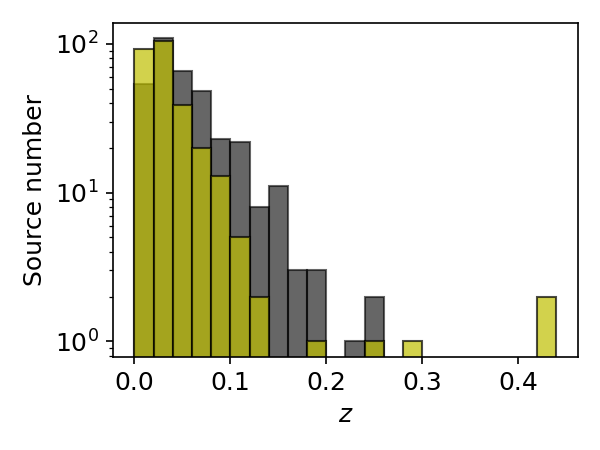}
\end{center}
\caption{Redshift distribution of Turin-SyCAT Seyfert galaxies. Type 1 in black, type 2 in yellow.}
\label{fig:z}
\end{figure}

\begin{figure}[ht]
\begin{center}
\includegraphics[width=8.cm]{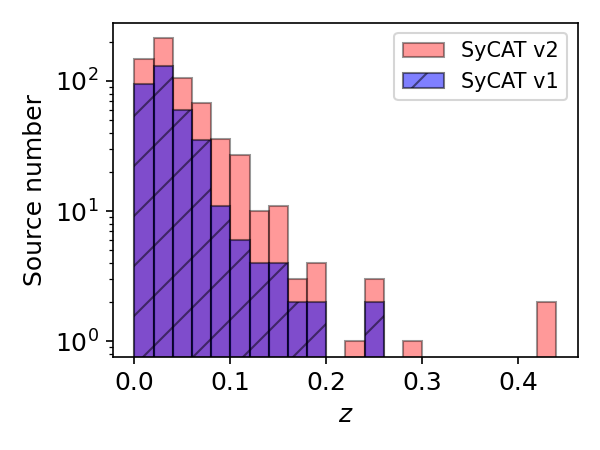}
\end{center}
\caption{Redshift distribution of all Seyfert galaxies from 1$^{st}$ release of the Turin-SyCAT compared to those listed in the 2$^{nd}$ release.}
\label{fig:z_sy1sy2_v1_v2}
\end{figure}

\begin{figure}[ht]
\begin{center}
\includegraphics[width=8.cm]{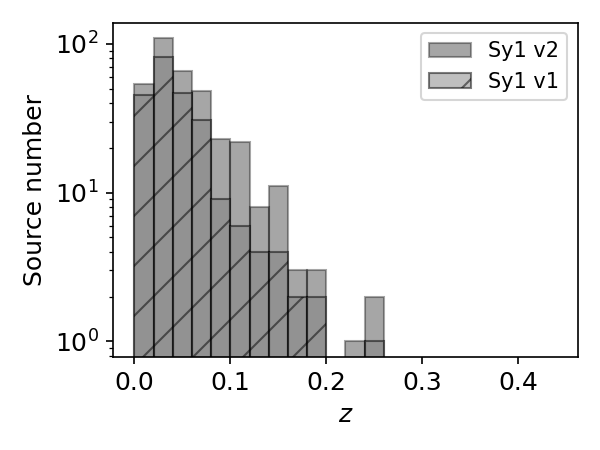}
\end{center}
\caption{Redshift distribution of Type 1 Seyfert galaxies from 1$^{st}$ release of the Turin-SyCAT compared to the presented 2$^{nd}$ release.}
\label{fig:z_sy1_v1_v2}
\end{figure}

\begin{figure}[ht]
\begin{center}
\includegraphics[width=8.cm]{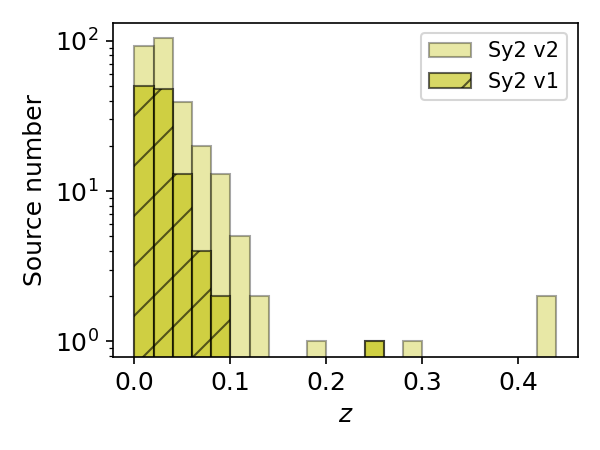}
\end{center}
\caption{Redshift distribution of Type 2 Seyfert galaxies from 1$^{st}$ release of the Turin-SyCAT compared to the presented 2$^{nd}$ release.}
\label{fig:z_sy2_v1_v2}
\end{figure}

In Figure~\ref{fig:z} we present the redshift distribution of Turin-SyCAT $2^{nd}$ release. The source number for both classes drastically drops after z $ > 0.2 $ as occurs for those listed in the first release, and the source with the highest redshift is SY2 J0304-3026 at 0.436. We compare the redshift distribution of all Seyfert galaxies (Figure~\ref{fig:z_sy1sy2_v1_v2}), only Type 1 Seyfert galaxies (Figure~\ref{fig:z_sy1_v1_v2}) and only Type 2 Seyfert galaxies (Figure~\ref{fig:z_sy2_v1_v2}) between the 1$^{st}$ release of the Turin-Sycat and it's presented 2$^{nd}$ release.

With respect to the previous Turin-SyCAT $1^{st}$ release, we modified the name of SY2 J2328+0330 in SY2 J2329+0331 having a WISE counterpart J232903.90+033159.9 and since a new Seyfert type 2 galaxy was associated with its mid-IR counterpart J232846.65+033041.1 thus being named as SY2 J2328+0330.

We list all sources included in the Turin-SyCAT $2^{nd}$ release in Table~\ref{tab:SyCAT} where we provide SyCAT $1^{st}$ release and SyCAT $2^{nd}$ release ids, SyCAT name, coordinates, spectroscopic redshifts, WISE counterpart, and 3PBC counterpart names, flux and flags to indicate if the source is also associated in 3PBC and BAT105 catalogs, and a flag to point out those added in this $2^{nd}$ release.

On the basis of the new Seyfert galaxies discovered here, we refined the connection between their hard X-ray and the mid-IR emission \citep{Assef2013}. This connection is related to the reprocessed radiation from the dust of all energy absorbed from the optical and UV wavelengths in the central engine of Seyfert galaxies (e.g. \cite{Elvis2009}). The high-energy emission measures an intrinsic radiated luminosity above $\sim$10 keV, while WISE 12 $\mu$m and 22 $\mu$m is related to the reprocessed radiation from the dust of all energy absorbed from the optical and UV wavelengths.

\begin{figure*}[ht]
\begin{center}
\includegraphics[height=5.8cm,width=8.cm]{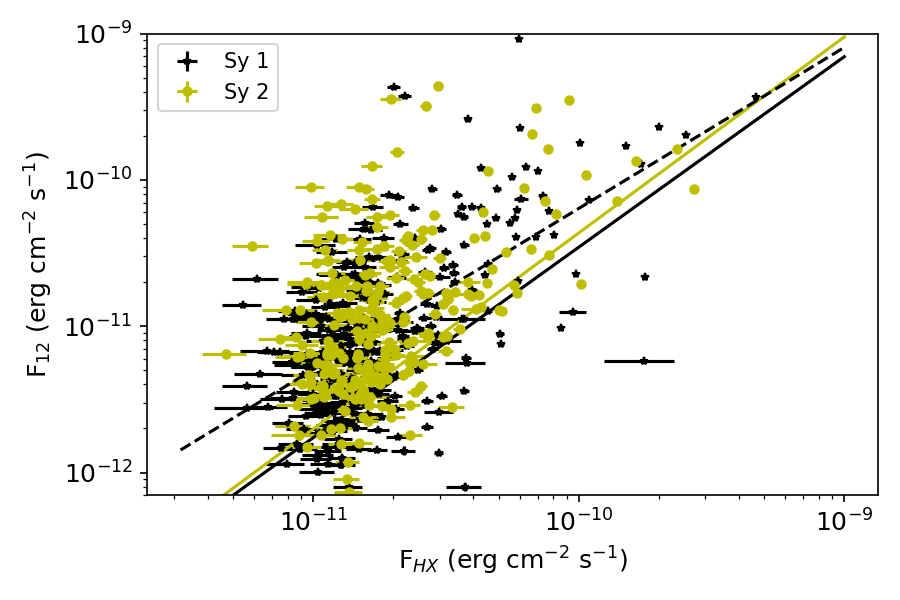}
\includegraphics[height=5.8cm,width=8.cm]{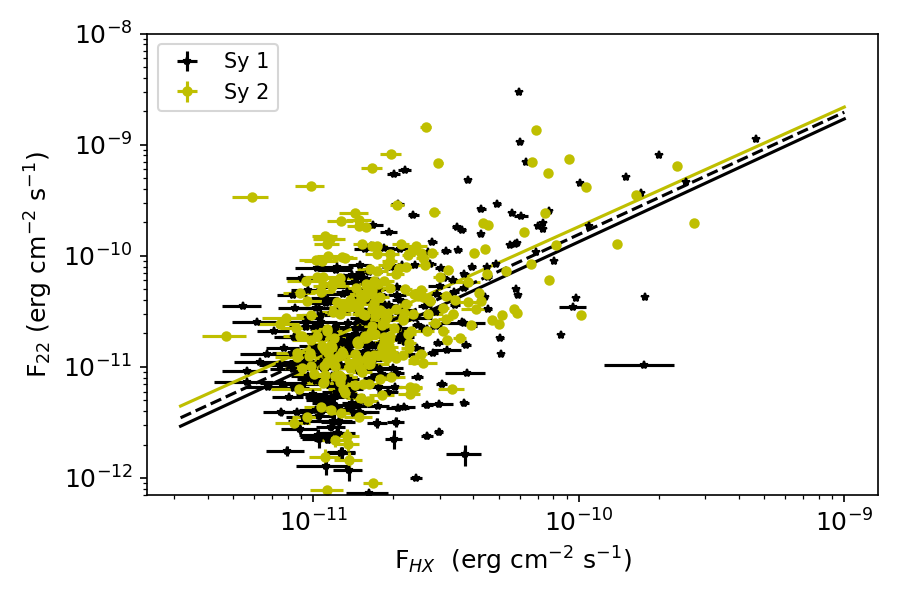}
\end{center}
\caption{ Mid-IR fluxes at 12 $\mu$m  (Left panel) and 22 $\mu$m (Right panel) as a function of hard X-rays flux. Regression lines were computed for the correlations between the W3 integrated flux and that in the hard X-ray band from the 3PBC, for both Seyfert 1 and 2 galaxies, marked in black and yellow, respectively. The dashed black line corresponds to the regression line computed for the whole sample while the straight black and yellow lines mark that for type 1 and type 2 Seyfert galaxies, respectively. The correlation coefficients are reported in Section~\ref{sec:section5}.
}
\label{fig:xconn}
\end{figure*}

Mid-IR fluxes show a significant correlation with the hard X-ray fluxes, similar to those highlighted using Seyfert galaxies listed in the Turin-SyCAT $1^{st}$ release, as shown in Figure~\ref{fig:xconn}. Comparing integrated fluxes as $F_{12}$ and $F_{HX}$ we found a linear correlation coefficient of 0.54 (correspondent to a slope of $ 1.09\pm0.10$ given the measured dispersion) for Seyfert 1 and 0.45 (slope of $1.20\pm0.16$) for Seyfert 2 galaxies, respectively. $F_{12}$ is the integrated flux at 12 microns derived from the WISE magnitude and $F_{HX}$ is the integrated hard X-ray flux in the 15-150 keV energy range both in units of erg cm$^{-2}$ s$^{-1}$. This is in agreement with results presented on the statistical analysis of Seyfert galaxies listed in the Turin-SyCAT $1^{st}$ release where we measured a correlation coefficient of 0.57, with a slope of 1.02 $\pm$ 0.10 and a coefficient of 0.52 (slope of 0.93 $\pm$ 0.16), for Seyfert 1 and 2 galaxies, respectively. On the other hand, also comparing mid-IR at lower frequencies with the hard X-ray flux (i.e., $F_{22}$ vs $F_{HX}$), where $F_{22}$ is the integrated flux at 22 microns derived from the WISE magnitude in units of erg cm$^{-2}$ s$^{-1}$, we found a correlation coefficient of 0.55 (with a slope of $1.11\pm0.10$) for type 1 Seyfert galaxies and 0.46 (slope of $1.08\pm0.17$) for type 2 Seyfert galaxies. 

Considering both classes together, since they show similar mid-IR to hard X-ray ratios, we found a correlation coefficient of 0.51 and a slope of $1.1\pm0.08$ for both hard X-ray flux $F_{HX}$ correlation with $F_{12}$ and $F_{22}$, all in agreement with previous results based on the Turin-SyCAT $1^{st}$ release.

\begin{figure*}[ht]
\begin{center}
\includegraphics[height=5.8cm,width=8.cm]{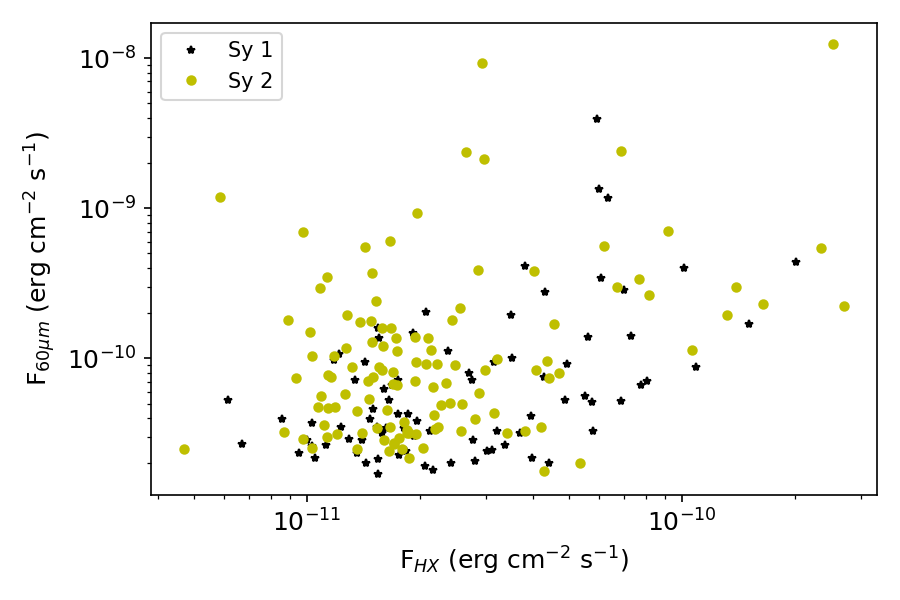}
\includegraphics[height=5.8cm,width=8.cm]{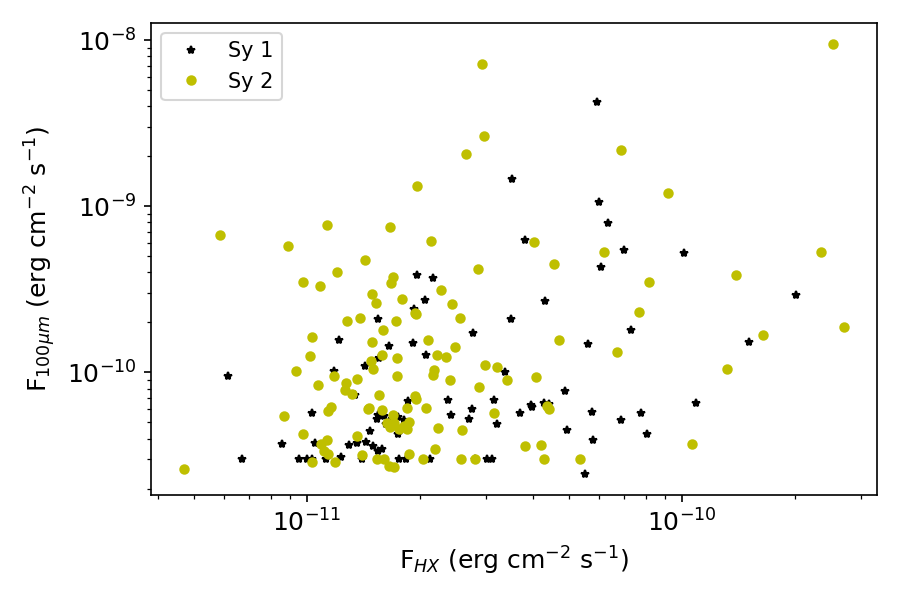}
\end{center}
\caption{Fluxes at 60 $\mu$m (left panel) and 100 $\mu$m (right panel) as a function of hard X-rays flux. Seyfert 1 and 2 galaxies are marked in black and yellow, respectively. No neat trend is evident between the two emissions.
}
\label{fig:iras}
\end{figure*}

We also cross-matched sources listed in Turin-SyCAT $2^{nd}$ release with the Point Source catalog of the Infrared Astronomical Satellite (\textit{IRAS})\footnote{\url{https://heasarc.gsfc.nasa.gov/W3Browse/iras/iraspsc.html}}, using the positional uncertainties reported therein. We obtained 67 new matches for a total of 216 Seyfert galaxies with an IRAS counterpart, 89 types 1 and 127 types 2 counterparts at both 60 $\mu$m and 100 $\mu$m, respectively. Then, as occurred in our previous analysis \citep{Herazo2022}, we also tested possible trends between the infrared fluxes, at 60 $\mu$m and at 100 $\mu$m, and the hard X-ray one. We found no clear correlation as evident in Figure~\ref{fig:iras} and again these results are in agreement with our previous findings based on Turin-SyCAT $1^{st}$ release. Moreover, we did not expect any correlation while inspecting trends between infrared and hard X-ray fluxes since the cold dust, mainly responsible for the emission at 60 $\mu$m and 100 $\mu$m is not significantly affected by the behavior of the central AGN but it is mainly linked to the star formation occurring in Seyfert galaxies \citep{Espinosa1987}.

The strict multi-frequency selection criteria that we used to select Turin-SyCAT sources allowed us to minimize the possible contamination of other source classes, thus strengthening our results. Thus we remind that we visually inspected all Turin-SyCAT galaxies' optical spectra, allowing us to measure their redshifts and establish their proper optical classification.

\begin{center}
\begin{sidewaystable}[] 
\tiny
\caption{The $2^{nd}$ version of the Turin-SyCAT catalog. Only the first 10 rows, the full catalog table is available in the online material.}
\label{tab:SyCAT}
\makebox[\textwidth]{
\begin{tabular}{rlllllllllll}
\hline
IDv2 & IDv1 & SyCAT & R.A. & Dec. & z & WISE & 3PBC & $F_{HX}$ & 3PBC flag & BAT105 flag & SyCAT v2 flag\\
 (1)   &  (2)   &  (3)   &  (4) & (5) & (6)   & (7) &  (8) & (9) & (10) & (11)& (12) \\
\noalign{\smallskip}
\hline 
\noalign{\smallskip}

\hline
  1 &  & SY2 J0000-0709 & 0.203216 & -7.153221 & 0.03748 & J000048.77-070911.6 & 3PBC J0000.9-0708 & 1.25E-11 $\pm$ 1.4E-12 & \checkmark & \checkmark & \checkmark\\
  2 &  & SY1 J0001-7657 & 0.441917 & -76.953972 & 0.05839 & J000146.08-765714.2 & 3PBC J0001.7-7659 & 1.09-11 $\pm$ 1.5E-12 & \checkmark & \checkmark & \checkmark\\
  3 & 1 & SY1 J0002+0322 & 0.6102917 & 3.352 & 0.025518 & J000226.41+032107.0 & 3PBC J0002.5+0322 & 1.39E-11 $\pm$ 1.9E-12 & \checkmark & \checkmark & --\\
  4 &  & SY2 J0003+2739 & 0.864283 & 27.654828 & 0.03969 & J000327.41+273917.0 & 3PBC J0003.4+2738 & 1.82E-11 $\pm$ 2.6E-12 & \checkmark & \checkmark & \checkmark\\
  5 & 2 & SY2 J0004+7020 & 1.00817 & 70.32175 & 0.096 & J000401.97+701918.2 & 3PBC J0004.0+7018 & 1.1E-11 $\pm$ 1.30E-12 & \checkmark & \checkmark & --\\
  6 & 3 & SY1 J0006+2013 & 1.58133 & 20.202917 & 0.025785 & J000619.53+201210.6 & 3PBC J0006.3+2012 & 1.76E-11 $\pm$ 1.40E-12 & \checkmark & \checkmark & --\\
  7 &  & SY1 J0017+0521 & 4.344167 & 5.352778 & 0.11 & J001722.71+052111.4 & 3PBC J0017.4+0519 & 8.69E-12 $\pm$ 1.5E-12 & \checkmark &  & \checkmark\\
  8 &  & SY2 J0021-1910 & 5.281417 & -19.168222 & 0.09558 & J002107.53-191005.4 & 3PBC J0021.1-1909 & 1.76E-11 $\pm$ 1.60E-12 & \checkmark & \checkmark & \checkmark\\
  9 & 4 & SY2 J0025+6821 & 6.38542 & 68.3622 & 0.012 & J002532.37+682144.9 & 3PBC J0025.5+6822 & 1.79E-11 $\pm$ 1.40E-12 & \checkmark & \checkmark & --\\
  10 & 5 & SY1 J0025-1859 & 6.4265 & -19.002917 & 0.24622 & J002542.34-190010.1 & 3PBC J0025.6-1859 & 1.05E-11 $\pm$ 1.60E-12 & \checkmark & -- & --\\
\hline
\noalign{\smallskip}
\hline
\end{tabular}
}
Column description:
(1) Unique catalog identified (ID) from SyCAT $2^{nd}$ version;
(2) Unique catalog identified (ID) from SyCAT $1^{st}$ version;
(3) SyCAT name;
(4) Right Ascencion J2000;
(5) Declination J2000;
(6) Redshift;
(7) name in WISE;
(8) name in 3PBC;
(9) Flux;
(10) flag if the source is in 3PBC;
(11) flag if the source is in  BAT105;
(11) flag if the source was added in SyCAT v2.
\end{sidewaystable}
\end{center}

\section{Summary, Conclusions and Future Perspectives}
\label{sec:section6}
The CXB is nowadays established to constitute mainly an integrated emission of discrete sources, primarily arising from AGNs \citep{Gilli2007}. Having a precise knowledge of the population and properties of various types of AGNs is thus crucial to improving our knowledge of the CXB.

In this work, we focus on the analysis of the 3PBC catalog \citep{Cusumano2010}, in particular focusing on the extragalactic source population, aiming also at discovering new Seyfert galaxies that can be included in the presented Turin-SyCAT $2^{nd}$ release. The 3PBC provides 1593 sources above signal to noise ratio of 3.8, approximately 57\,\% sources appear to have a clear extragalactic origin while 19\,\% belong to our Milky Way, and the remaining 24\,\% are yet unknown. Results of our multifrequency investigation are also based on those recently found for the 105 month \textit{Swift}-BAT catalogue \citep{Oh2018} and the \textit{INTEGRAL} IBIS hard X-ray survey in energy range 17-100 keV \citep{Bird2016}.
For comparison, the original release of the 3PBC catalog listed 521 Seyfert galaxies, 109 blazars, 362 unclassified sources, and 244 unidentified sources, all classified according to our classification scheme while in the refined version presented here there are 593 Seyfert galaxies, 129 blazars, 199 unclassified sources, and 218 unidentified sources. It is worth highlighting that, on the basis of our classification criteria, we indicated 98 sources as unclassified, even though they had an assigned class in the original 3PBC catalog, due to a lack of information. All details about how we interpreted 3PBC original classes according to our classification scheme are reported in Appendix \ref{app:3PBC_reclassification} and Table \ref{table:3PBC_reclassification}.

Thanks to our analysis we (i) developed a multifrequency classification scheme for hard X-ray sources, that can be later adopted also to investigate different high-energy surveys, (ii) investigate the main properties of sources populating the extragalactic hard X-ray sky and finally, extract other Seyfert galaxies now included in the $2^{nd}$ release of the Turin-SyCAT catalog presented here.

We worked with the 1593 sources of the 3PBC catalog, comparing them with various other catalogs mentioned in the paper and adopting the following classification scheme criteria. Firstly, we checked if the 3PBC source has an assigned counterpart, if not, we performed multifrequency crossmatching analyses across the available literature to search for counterparts. Sources without counterparts were assigned with \textit{unidentified} category. Those which were found with counterparts, together with sources already having a counterpart in the 3PBC catalog, were further inspected with multifrequency analyses. Sources lacking sufficient information to assign their class were put to \textit{unclassified} category, the rest to \textit{classified} category. We further distinguish the classified sources into Galactic and extragalactic and we purely focus on the extragalactic sources in this work.

Results obtained from our analysis can be outlined as follows.
\begin{enumerate}
\item The final revised 3PBC catalog we present in this study lists 1176 classified, 820 extragalactic, and 356 Galactic ones, 218 unidentified and 199 unclassified sources, respectively. 
The original version of the 3PBC catalog listed 244 unidentified and 362 unclassified sources counted according to our classification scheme (see Appendix for more details). We improved the fraction of 15.3\% unidentified sources to 13.7\% (from 244 to 218 sources) and the fraction of 22.7\% unclassified sources to 12.5\% (from 362 to 199 sources). It is important to highlight that 98 sources were classified in the original 3PBC catalog, but they were indeed listed as unclassified according to our refined analysis since they lacked multifrequency information.
\item The hard X-ray sky is mainly populated by nearby AGNs, where the two largest known populations of associated AGNs are: Seyfert galaxies ($\sim 79\%$) and blazars ($\sim 17\%$).
\item We report the trends between the hard X-ray and the gamma-ray emissions of those blazars that are also listed in the 4FGL as expected by the models widely adopted to explain their broadband SED. 
\item In the presented $2^{nd}$ release of the Turin-SyCAT, we list 633 Seyfert galaxies, 282 new ones added here thus correspondent to increase their number by $\sim$80\,\% with respect to its $1^{st}$ release.
\item We updated the statistical analysis carried out comparing the hard X-ray and the IR emissions of Seyfert galaxies.  All results obtained are in agreement with those previously found even if now the analysis appears more robust since it was performed with a sample of Seyfert galaxies increased by $\sim$80\,\% with respect to the 1$^{st}$ release of the Turin-SyCAT.
\end{enumerate}

Finally, we already checked the presence of SWIFT observations carried out using the X-ray telescope on board for the sample of unidentified hard X-ray sources and we found that more than 95\,\% of them have at least a few ksec exposure time available. Thus the next step of the present analysis will be to search for the potential soft X-ray counterpart of these 3PBC unidentified sources to obtain their precise position necessary to carry out optical spectroscopic campaigns aimed at identifying the whole sky seen between 15 and 150 keV.

\begin{table*}[h]
 \caption[]{\label{table:catalogs_table}Table of catalogs used in the cross-matching analysis.}
\begin{tabular}{ccc}
 \hline \hline
  Acronym &
  Catalogue Name &
  Reference \\
\hline
\hline
4FGL-DR2    & The second release of the fourth \textit{Fermi}-LAT catalog \\
            & of $\gamma$-ray sources                                        & 1\\
3PBC        & The 3$^{rd}$ Palermo \textit{Swift}-BAT Hard X-ray catalog               & 2\\
BAT105      & The 105-month \textit{Swift}-BAT catalog             & 3\\
\textit{INTEGRAL}    & The IBIS soft gamma-ray sky after 1000 \textit{INTEGRAL} orbits & 4\\
Homa-BZCAT  & 5$^{th}$ edition of Roma-BZCAT catalog of blazars                 & 5\\
3CR         & The Revised Third Cambridge catalog                          & 6\\
4C          & The Fourth Cambridge Survey                                    & 7, 8\\
SyCAT       & The Turin-SyCAT catalog                                      & 9\\
CVcat       & The Catalog and Atlas of Cataclysmic Variables                 & 10\\
SNRcat      & The Catalog of Galactic Supernovae Remnants                    & 11\\
hmxb        & The 4$^{th}$ edition of the catalog of High mass X-ray binaries \\
            & in the Galaxy                                                  & 12\\
lmxb        & The 4$^{th}$ edition of the catalog of Low mass X-ray binaries \\ 
            & in the Galaxy and Magellanic Clouds                            & 13\\
Rlmxb      & The 7$^{th}$ edition of the catalog of cataclysmic binaries, \\ 
            & low mass X-ray binaries and related objects                    & 14\\
ANTF      & The Australian Telescope National Facility Pulsar Catalog      & 15\\
Abellcat    & Abell catalog of rich galaxy clusters                        & 16\\
\hline
\end{tabular}
\tablebib{(1)~\citet{Ballet2020};
          (2) \citet{Cusumano2010}; 
          (3) \citet{Oh2018}; 
          (4) \citet{Bird2016};
          (5) \citet{Massaro2015}; 
          (6) \citet{Spinrad1985}; 
          (7) \citet{Pilkington1965};
          (8) \citet{Gower1967};
          (9) \citet{Herazo2022};
          (10) \citet{Downes2005}; 
          (11) \citet{Green2017};
          (12) \citet{Liu2006};
          (13) \citet{Liu2007};
          (14) \citet{Ritter2003};
          (15) \citet{Manchester2005};
          (16) \citet{Abell1989}.
}
\end{table*}

\begin{acknowledgements}
We thank the anonymous referee for useful comments that led to improvements in the paper. M. K. and N. W. are supported by the GACR grant 21-13491X. E. B. acknowledges NASA grant 80NSSC21K0653. M. K. was supported by the Italian Government Scholarship issued by the Italian MAECI. VC acknowledges support from CONACyT research grants
280789 (Mexico). F. M. wishes to thank Dr. G. Cusumano for introducing him to the Palermo BAT Catalog project. We would like to thank A. Capetti for his work done on the 1$^{st}$ version of the Turin-SyCAT, which was relevant for this work. This investigation is supported by the National Aeronautics and Space Administration (NASA) grants GO0-21110X, GO1-22087X, and GO1-22112A. This research has made use of the NASA/IPAC Infrared Science Archive, which is funded by the National Aeronautics and Space Administration and operated by the California Institute of Technology. Funding for the Sloan Digital Sky Survey IV has been provided by the Alfred P. Sloan Foundation, the U.S. Department of Energy Office of Science, and the Participating Institutions. SDSS-IV acknowledges support and resources from the Center for High-Performance Computing at the University of Utah. The SDSS website is www.sdss.org. SDSS-IV is managed by the Astrophysical Research Consortium for the Participating Institutions of the SDSS Collaboration including the Brazilian Participation Group, the Carnegie Institution for Science, Carnegie Mellon University, Center for Astrophysics | Harvard \& Smithsonian, the Chilean Participation Group, the French Participation Group, Instituto de Astrof\'isica de Canarias, The Johns Hopkins University, Kavli Institute for the Physics and Mathematics of the Universe (IPMU) / University of Tokyo, the Korean Participation Group, Lawrence Berkeley National Laboratory, Leibniz Institut f\"ur Astrophysik Potsdam (AIP),  Max-Planck-Institut f\"ur Astronomie (MPIA Heidelberg),Max-Planck-Institut f\"ur Astrophysik (MPA Garching), Max-Planck-Institut f\"ur Extraterrestrische Physik (MPE), National Astronomical Observatories of China, New Mexico State University, New York University, University of Notre Dame, Observat\'ario Nacional / MCTI, The Ohio State University, Pennsylvania State University, Shanghai Astronomical Observatory, United Kingdom Participation Group, Universidad Nacional Aut\'onoma de M\'exico, University of Arizona, University of Colorado Boulder, University of Oxford, University of Portsmouth, University of Utah, University of Virginia, University of Washington, University of Wisconsin, Vanderbilt University, and Yale University. The Pan-STARRS1 Surveys (PS1) and the PS1 public science archive have been made possible through contributions by the Institute for Astronomy, the University of Hawaii, the Pan-STARRS Project Office, the Max-Planck Society, and its participating institutes, the Max Planck Institute for Astronomy, Heidelberg and the Max Planck Institute for Extraterrestrial Physics, Garching, The Johns Hopkins University, Durham University, the University of Edinburgh, the Queen's University Belfast, the Harvard-Smithsonian Center for Astrophysics, the Las Cumbres Observatory Global Telescope Network Incorporated, the National Central University of Taiwan, the Space Telescope Science Institute, the National Aeronautics and Space Administration under Grant No. NNX08AR22G was issued through the Planetary Science Division of the NASA Science Mission Directorate, the National Science Foundation Grant No. AST–1238877, the University of Maryland, Eotvos Lorand University (ELTE), the Los Alamos National Laboratory, and the Gordon and Betty Moore Foundation. This publication makes use of data products from the Wide-field Infrared Survey Explorer, which is a joint project of the University of California, Los Angeles, and the Jet Propulsion Laboratory/California Institute of Technology, funded by the National Aeronautics and Space Administration. TOPCAT and STILTS astronomical software \citep{Taylor2005} were used for the preparation and manipulation of the tabular data and the images.

\end{acknowledgements}

\bibliographystyle{aa} 
\bibliography{aanda}

\begin{appendix} 

\section{Re-classification of the original 3PBC based on our classification scheme}
\label{app:3PBC_reclassification}

We compare our refined classification for the counterparts of 3PBC sources with those previously assigned in the original catalog. To perform this, we first ``translate'' the original 3PBC classes into our classification scheme according to the following criteria. For the Galactic objects we consider (i) all sources classified in the original 3PBC as HXB, LXB, XB, XB*, V* being indicated as ``bin'' (i.e., binary systems); (ii) those previously labeled as AM*, CV, CV*, DN*, DQ*, EB*, NL*, No currently belong to the ``cv'' class (i.e., cataclysmic variables); while (iii) those few classified as Psr are all ``psr'', SNR is ``snr'' and PN simply ``pn'' (i.e., pulsars, supernova remnants, and planetary nebulae, respectively). On the other hand, for the extragalactic classes: (i) sources labeled as BLA and BZC in the original 3PBC are indicated as ``blz'' (i.e., blazars); (ii) Sy*, SyG, Sy1, Sy2 are all classified as ``sey'' according to our scheme (i.e., Seyfert galaxies) while (iii) QSO are ``qso'' and rG is ``rdg'' (i.e., being quasars and radio galaxies respectively) and then (iv) objects classified as LIN and ClG are now indicated as ``liner'' and ``clu'', respectively being LINERs and galaxy clusters. The remaining handful of sources had the same classification in both lists, for example, the Galactic center.

For the unknown sources, we considered those having a question mark in the classification label, remarking that this could be unsettled, as well as those indicated as AGN, BRT, EmG, G, GiC, GiP, IG, IR, X, gam and Rad, all ``unc''. This is because even if for example the associated counterpart in the original 3PBC is recognized as an infrared or an X-ray source, or a simple AGN, this does not provide us precise information about its nature and its hard X-ray emission. Finally, 3PBC sources originally lacking an assigned counterpart were all labeled as ``uhx'', being unidentified hard X-ray sources.

In Table \ref{table:3PBC_reclassification} we report the (i) 3PBC name, (ii) the original 3PBC classification, (iii) the new label corresponding to our new classification scheme but assigned on the basis of the previous criteria and on the information available before our refined analysis and (iv) our new classification based on the multifrequency analysis carried out here. This allowed us to compare previous and new associations and classifications to obtain an estimate of the improvements achieved.

We found that, according to our classification scheme, the 3PBC catalog presented 521 Seyfert galaxies, 109 blazars, 362 unclassified sources, and 244 unidentified sources. In our revised version of the 3PBC, we present 593 Seyfert galaxies, 129 blazars, 199 unclassified sources, and 218 unidentified sources. It is important to highlight, that due to our classification criteria, we not only classified some of the yet unclassified or unidentified sources, but we also re-classified 98 sources that had a classification class in the original 3PBC, as unclassified according to our classification criteria. This was done in cases of a lack of information in the literature, e.g. if we could not find an optical spectrum or luminosities, etc.

\begin{table}[h]
\caption{The comparison between the original classes assigned in the 3PBC, how they are interpreted according to our new scheme and that obtained thanks to our refined analysis, for all 3PBC sources. For each source, we report the following columns: (i) the 3PBC name; (ii) the original class; (iii) the class interpreted according to our scheme; (iv) the class assigned in our refined analysis.} 
\label{table:3PBC_reclassification}      
\begin{tabular}{ccccc}
 \hline \hline
 name\_3PBC & class3PBC & reclass & class & subclass\\
\hline
J0000.9-0708 & X & unc & sey & sy2 \\
J0001.7-7659 & G & unc & sey & sy1 \\
J0002.5+0322 & Sy1 & sey & sey & sy1 \\
J0003.4+2738 & G & unc & sey & sy2 \\
J0004.0+7018 & AG? & unc & sey & sy2 \\
\hline
\end{tabular}
\end{table}


\end{appendix}

\end{document}